\documentclass[%
 reprint,
superscriptaddress,
 amsmath,amssymb,
 aps,
 pra,
]{revtex4-1}

\usepackage{CJK}

\usepackage{graphicx}
\usepackage{dcolumn}
\usepackage{bm}
\usepackage{dcolumn}
\newcolumntype{d}[1]{D{.}{.}{#1}}
\renewcommand{\vec}[1]{\boldsymbol{#1}} 

\usepackage{color}
\definecolor{Blue}{rgb}{0.3,0.3,0.9}
\definecolor{Red}{rgb}{0.9,0.3,0.3}
\definecolor{Green}{rgb}{0.3,0.6,0.3}
\newcommand{\revision}[1]{#1}

\newif\ifNOSUP \NOSUPfalse

\begin{document}
\begin{CJK*}{UTF8}{gbsn}


\title{Unconventional delocalization in a family of 3D Lieb lattices
}

\author{Jie Liu (刘洁)}
\email{liujie@smail.xtu.edu.cn}
\affiliation{School of Physics and Optoelectronics, Xiangtan University, Xiangtan 411105, China}

\author{Carlo Danieli}
\email{carlo.danieli@roma1.infn.it}
\altaffiliation[Current Address: ]{Department of Physics, University of Sapienza, Piazzale Aldo Moro 5, 00185 Rome, Italy}
\affiliation{Max Planck Institute for the Physics of Complex Systems, Dresden D-01187, Germany}

\author{Jianxin Zhong (钟建新)}
\email{jxzhong@xtu.edu.cn}
\affiliation{School of Physics and Optoelectronics, Xiangtan University, Xiangtan 411105, China}

\author{Rudolf A. R\"omer}
\email{r.roemer@warwick.ac.uk}
\affiliation{School of Physics and Optoelectronics, Xiangtan University, Xiangtan 411105, China}
\affiliation{Department of Physics, University of Warwick, Coventry, CV4 7AL, United Kingdom}

\date{\today}

\begin{abstract}
Uncorrelated disorder in generalized 3D Lieb models gives rise to the existence of bounded mobility edges, destroys the macroscopic degeneracy of the flat bands and breaks their compactly-localized states. 
We now introduce a mix of order and disorder such that this degeneracy remains and the compactly-localized states are preserved. We obtain the energy-disorder phase diagrams and identify mobility edges. Intriguingly, for large disorder the survival of the compactly-localized states induces the existence of delocalized eigenstates close to the original flat band energies -- yielding seemingly divergent mobility edges. For small disorder, however, a change from extended to localized behavior can be found upon decreasing disorder --- leading to an unconventional ``inverse Anderson'' behavior. We show that transfer matrix methods, computing the localization lengths, as well as sparse-matrix diagonalization, using spectral gap-ratio energy-level statistics, are in excellent quantitative agreement. The preservation of the compactly-localized states even in the presence of this disorder might be useful for envisaged storage applications.
\end{abstract}

\maketitle
\end{CJK*}

\section{\label{sec:intro}Introduction}

The phenomenon of wave localization in disordered lattices has attracted a lot of attention in the condensed matter community since it was first predicted in 1958~\cite{Anderson1958c} for uncorrelated random potentials. 
The resulting localization properties induced by the disorder can strongly depend on the lattice dimensionality, the type of lattice geometry considered, as well as the nature of the potential considered \cite{Krameri1993,2003AndersonRamifications,Evers2008}. 
Indeed, if in a $2D$ square lattices with uncorrelated disorder all eigenstates are exponentially localized for any disorder strength~\cite{Abrahams1979ScalingDimensions}, this is no longer true in a $3D$ cubic lattice where an energy-dependent transition from delocalized to localized eigenstates is induced only after reaching a critical disorder strength~\cite{Bulka1985}. 
Likewise, a transition from delocalized to localized phase may also occur in $1D$ chains when correlated disordered potentials are considered~\cite{Aubry1980AnalyticityLattices,Izrailev1999}. 

Spatial disorder, however, is not the only ingredient that can lead to wave localization phenomena in lattices. In translationally invariant networks, one of the most intensely studied frameworks for eigenstates localization is the case of flat band lattices -- \emph{ i.e.}~networks where destructive interference results in families of macroscopically degenerate single-particle eigenstates localized within a finite number of lattice sites~\cite{Derzhko2015a,Leykam2018,Leykam2018c}. 
These states, called \emph{compact localized states} (CLS), form a non-dispersive (hence, \emph{flat}) Bloch band $E_j({\bf k}) = \text{const.}$ in the energy spectrum which is independent on the momentum ${\bf k}$. 
First introduced to analytically study ferromagnetic ground states in many-body systems~\cite{Sutherland1986b,Lieb1989TwoModel}, flat band models have since been used to study a plethora of physical phenomena, from 
the fractional quantum Hall effect~\cite{Tang2011,Neupert2011,Sun2011NearlyTopology}, to spin liquids~\cite{Savary2017,Balents2010},
ferromagnetism~\cite{Mielke1991a,Tasaki1992a,Mielke1993FerromagnetismModel,Ramirez1994}, disorder-free many-body localization~\cite{Danieli2020,Kuno2020a},
superfluidity and superconductivity~\cite{Miyahara2007BCSLattice,Julku2016,Kopnin2011,Peotta2015,Tovmasyan2018PreformedBands,Mondaini2018,Aoki2020TheoreticalSuperconductivity}, among others. 
Furthermore, flat band systems have also been experimentally realized in a variety of diverse settings, such as electronic systems~\cite{Abilio1999},
ultracold atomic systems~\cite{Shen2010,Goldman2011,Apaja2010} and 
photonic systems \cite{Mukherjee2015a,Vicencio2015a,Guzman-Silva2014,Diebel2016,Taie2015,Nixon2013}. 

The CLS have been discussed as potential candidates for information storage applications \cite{Rontgen2019}. However, they are typically sensitive to perturbations. Uncorrelated onsite disorder in most cases lifts the existence of CLS irrespective of the disorder strength and induces wave localization in flat band lattices~\cite{Chalker2010a,Leykam2013,Flach2014a,Leykam2017,Bilitewski2018,Shukla2018a,Mao2020b,Cadez2021}. 
In certain cases, however, local symmetries within flat band lattices suggest local correlations in the onsite disorder which result in anomalous localization features -- as shown in Refs.~\cite{Bodyfelt2014,Danieli2015} for disorder and quasiperiodic potentials in $1D$ and $2D$ sample lattices. 

In this work we study the impact of local ordering correlations in a family of $3D$ extended Lieb lattices. These lattice systems, in presence of uncorrelated spatial disorder, exhibit energy-dependent transitions between localized to delocalized phase~\cite{Liu2020a}. 
By exploiting local symmetry in the family of Lieb lattices, we introduce a mix of correlated order and disorder within the lattice. This mix of local order and disorder preserves the existence of the degenerate CLS and induces an effective projection of the non-degenerate states onto the CLS~\cite{Chalker2010a}. 
The projection yields the existence of delocalized states existing mostly within the locally ordered sub-lattice of the systems spanned by the CLS, whose energies lie closer to the macroscopic degeneracy as the strength of the disorder increases. Ultimately, the persistence of these extended states results in a divergent profile of the mobility edge separating delocalized and localized phases, unlike what was found in Ref.~\cite{Liu2020a} for uncorrelated disorder.   
Furthermore, we observe that this correlated ordering in the regime of weak disorder induces an ``inverse" change from localized to delocalized eigenstates for energies close to the macroscopic degeneracies. 

The paper is structured as following: 
in Sec.~\ref{sec:model} we introduce the extended Lieb lattices called  $\mathcal{L}_3(n)$ and review the numerical methods employed, 
while in Sec.~\ref{sec:res} we present our results, separating between the standard $3D$ Lieb lattice  $\mathcal{L}_3(1)$ in Sec.~\ref{sec:L31} and its generalized version $\mathcal{L}_3(2)$ and beyond  in Sec.~\ref{sec:L32}. We conclude in Sec.\ \ref{sec:conclusions}.

\section{\label{sec:model}Models and Methods}

\subsection{The extended Lieb Models in 3D}

\begin{figure*}[tb]
    \centering
    \includegraphics[width=1.95\columnwidth]{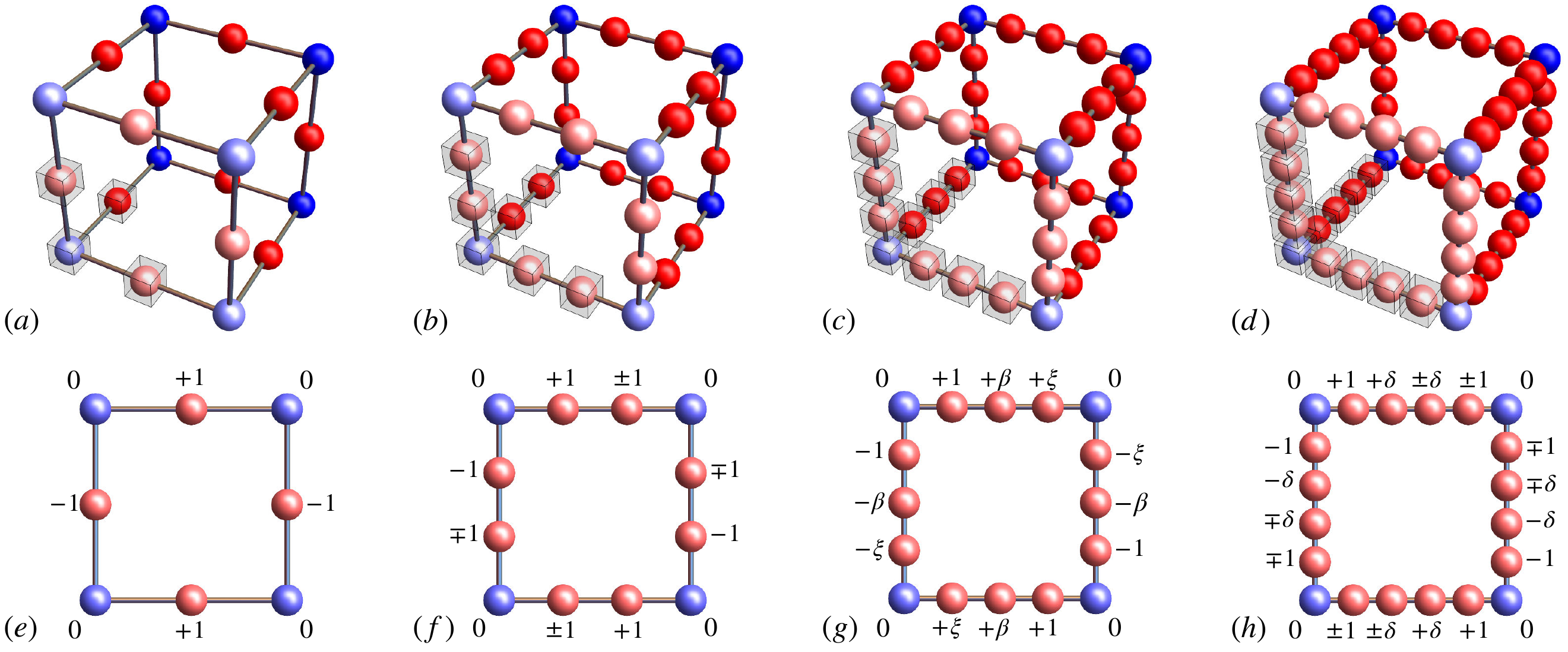}
    \caption{(a-d) Schematic representation of the first four cases of the lattice family $\mathcal{L}_3(n)$ -- namely (a) $\mathcal{L}_3(1)$, (b) $\mathcal{L}_3(2)$, (c) $\mathcal{L}_3(3)$ and (d) $\mathcal{L}_3(4)$. Blue spheres denote the cube sites while red spheres denote the Lieb sites. The latter are without disorder to retain the degenerate compact states in our model. \revision{Sites comprising the unit cell in the bottom left corner of each lattice are enclosed in a (gray) cube. The dark lines between sites} are guides to the eye and indicate the hopping profiles. The light blue and \revision{light} red sites denote the front plaquettes.
    (e-h) Plaquettes with their CLS indicated by their (unnormalized) wave function amplitudes \revision{$| \vec{X} \rangle$}. In (e-f) the flat bands are (e) $E=0$ for $n=1$ and (f) $E=\pm1$ for $n=2$. 
    In (g), for $n=3$ the three flat bands are $E=\beta = 0,\pm\sqrt{2}$, with $\xi=+1$ for $\beta = \pm\sqrt{2}$ and $\xi=-1$ for $\beta = 0$. 
    In (h), for $n=4$ the four flat bands are $E=\pm \delta$, with $\delta =\frac{1}{2}(1\pm\sqrt{5}$). 
    }
    \label{fig:Lieb_schematic}
\end{figure*}

We consider a parametric family of three-dimensional Lieb lattices labeled $\mathcal{L}_3(n)$, $n=1, 2, \ldots$, and defined by the Hamiltonian
%
\begin{equation}
    H=\sum_{\vec{X}} \varepsilon_{\vec{X}} |\vec{X}\rangle \langle \vec{X}|
    - \sum_{\vec{X}\neq \vec{Y}}t_{\vec{X}\vec{Y}}|\vec{X}\rangle \langle \vec{Y}|\ .
\label{Equ:def1} 
\end{equation} 
Here, the set of $|\vec{X}\rangle$ indicates the orthonormal Wannier states corresponding to electrons located at sites $\vec{X}=(x,y,z)$ of the Lieb lattices and $\varepsilon_{\vec{X}}$ is the onsite potential \cite{Liu2020a}. As usual, we set the hopping integrals $t_{\vec{X},\vec{Y}} \equiv 1$ for nearest-neighbor sites $\vec{X}$ and $\vec{Y}$ and $t_{\vec{X},\vec{Y}} \equiv 0$ otherwise. 
The integer parameter $n$ enumerates the added number of sites between two sites located at the ``cubic" vertexes of the lattices as shown in Fig.~\ref{fig:Lieb_schematic} for $n=1$ to $4$. 
We denote those sites sitting on the vertexes of the lattices as the \emph{cube} sites (colored in blue in Fig.~\ref{fig:Lieb_schematic}) while those sites located between two neighboring cube sites are called the \emph{Lieb} sites (colored in light red in Fig.~\ref{fig:Lieb_schematic}). Hence $3n+1$ is the total number of sites per unit cell, resulting in $3n+1$ bands.
%
\revision{Notably, for any $n$, the corresponding lattices $\mathcal{L}_3(n)$ have $n$ double-degenerate flat bands (namely, all flatbands are counted twice). Thus, for any $n$, there exist $n$-families of macroscopically degenerate compactly localized states (CLS), all of which have strictly non-zero amplitude in the Lieb sites enclosed within each 2D square plaquette of the lattice -- as shown in Fig.~\ref{fig:Lieb_schematic}(e-h) for $n=1$ to $4$. Further details of the CLS on generalized Lieb lattices $\mathcal{L}_3(n)$ are given in the supplemental material \cite{supp}.
}

We consider in Eq.~\eqref{Equ:def1} \emph{locally correlated potentials} $ \varepsilon_{\vec{X}}$ which
neither destroy the existence of CLS nor renormalize their degeneracy while simultaneously offering the possibility of localization for non-CLS states. To ensure this for any number $n$ of Lieb sites, the simplest choice is to set the onsite potential of Lieb sites $\varepsilon_{\vec{X}}^{(L)}$ constant, \emph{i.e.}\ $\varepsilon_{\vec{X}}^{(L)}\equiv 0$, while introducing a spatially varying disorder potential on the cube sites $\varepsilon_{\vec{X}}^{(c)}$ via uncorrelated uniform random numbers with disorder strength $W$ such that $\varepsilon_{\vec{X}}^{(c)}\in \left[-\frac{W}{2}, \frac{W}{2}\right]$.
Note that in this setup of mixed order and disorder, the standard 3D Anderson model of localization \cite{Anderson1958c} can be recovered for $n=0$. 
  
\subsection{Transfer matrix-based measures of localization}

To study the emerging localization features due to such locally correlated potentials, we combine diverse numerical methods applied to finite versions of these 3D lattices. 
We compute the reduced localization length $\Lambda_M$ of a wave-function by the usual transfer-matrix method (TMM) \cite{MacKinnon1983a,Liu2020a}. 
In brief, the method considers electrons transferring, according to the single particle, stationary Schr\"{o}dinger equation, along a quasi-1D bar with fixed transversal square cross section of $M^2$ unit-cells for given ${\cal L}_3(n)$ via highly optimized matrix-vector calculations. One iteratively obtains converging estimates of self-averaged localization length $\lambda_M(E,W)$, with $\Lambda_M= \lambda_M(E,W)/M$ the dimensionless, \emph{reduced} localization length, when the number of electron transfers $\tilde{M}$, \emph{i.e.}\ the number of matrix-vector calculations in the longitudinal direction, is typically $\tilde{M} > 10^7$--$10^9$ such that $\tilde{M} \gg M$ along the bar~\cite{Note2,Romer2022NumericalLocalization}.

%
 
A system-size-independent intersection point of the $\Lambda_M(W)$ curves obtained for different bar widths $M$ at a given energy $E$ (or, alternatively, versus $E$ at a given disorder $W$) can indicate a critical disorder $W_c$ (respectively, the critical energy $E_c$), at least for large enough $M$.
Such critical values mark a transition between a metallic/extended/delocalized regime, where $\Lambda_M$ monotonically increases as $M$ grows, and an insulating/localized regime, where $\Lambda_M$ monotonically decreases as $M$ grows. Expecting the metal-insulator transition to be a second order phase transition~\cite{Krameri1993,Belitz1994a,Evers2008}, we can then extract a critical exponent $\nu$ characterizing the divergence of the correlation length $\xi(W)\sim |W-W_c|^{-\nu}$ (respectively, $\xi(E)\sim |E-E_c|^{-\nu}$) via finite-size scaling (FSS) -- under the assumption of single parameter scaling via $\Lambda_M(E,W,M)=f\left(\xi(E,W)/M\right)$ ~\cite{Krameri1993,Slevin1999b}. In principle, this $\nu$ should determine the universality class of the model. For the standard Anderson model with $n=0$, one finds $\nu=1.590(6)$ as well as $W_c=16.530(3)$ \cite{Rodriguez2011MultifractalTransition}.
A more detailed technical description of these methods as applied to Lieb lattices is given in Ref.~\cite{Liu2020a} and in the papers cited therein.

\subsection{Spectral measures of localization}

Initial results for the ${\cal L}_3(n)$ models obtained from the TMM calculations appear, at least at first glance, somewhat unexpected and hint towards a surprisingly rich phase structure when compared to the well-known Anderson behaviour.
We therefore proceed to also compute the density of states (DOS) and various energy level-ratio statistics via diagonalization of the Hamiltonian in Eq.\ \eqref{Equ:def1}. 
We compute spectra of the models via (i) exact diagonalization \cite{Note3} 
for the complete spectrum with typically ${\cal O}(10^2)$ potential configurations and (ii) sparse-matrix diagonalization \cite{Bollhofer2007JADAMILU:Matrices} for selected energy ranges in the spectrum  with typically ${\cal O}(10^4)$ potential configurations. Each such diagonalization routines is applied to a cubic section of a $\mathcal{L}_3(n)$ lattice with periodic boundary conditions. With $N$ denoting the number of unit-cells, we then have a total of $L = (3n+1) N^3$ sites. We emphasize that the given typical numbers of potential realization have to be realized for each $L$ and each $W$, resulting in sizable computational run-time requirements even on modern taskfarm installations. Furthermore, we took care to make all such configurations use independent random numbers --- otherwise results would show inconsistencies in the error estimates.

With eigenenergies $E_i$, the DOS is simply given as a suitable histogram of $E_i$ values. For the energy level-ratio statistics, we start with 
the \emph{adjacent gap ratio} $r_{i} = \min(s_{i}, s_{i+1})/\max(s_{i}, s_{i+1})$ with $s_{i} = E_i - E_{i-1}$ which can discriminate between extended and localized phases \cite{Oganesyan2007c}. In the extended phase, the $r$-values follow the gap-ratio distribution $P(r)$ of the Gaussian orthogonal matrix ensemble (GOE) with numerically determined mean value $\langle r\rangle = \int_{0}^{1} r P(r) \text{d} r = 0.5295$~\cite{Oganesyan2007c} or analytical surmise $\langle r\rangle_\text{Sur} = 4-2\sqrt{3}\approx 0.53590$~\cite{Atas2013b}. In the localized phase, the $r$-values follow the $P(r)$ for a Poisson random number distribution, with mean value $\langle r\rangle_\text{Poi} =2\ln{2}-1\approx 0.386$. 
Next, we compute from the spetra a more recent measure introduced and used in Refs.~\cite{Sa2020,Luo2021a} by defining the \emph{extended gap ratio} $|z_i|= |E_i-E_{NN}|/|E_i-E_{NNN}|$ and $E_{NN}$ and $E_{NNN}$ the nearest and the second-nearest eigenenergies to $E_i$, respectively. In this case, the mean value $\langle |z|\rangle$ ranges between  $\langle |z|\rangle_\text{ext} = 0.5687(1)$ (extended, \emph{i.e.}\ GOE matrices) and $\langle |z|\rangle_\text{loc} = 0.5000(1)$ (localized, \emph{i.e.}\ Poisson matrices) \cite{Note4}.
%
For both measures, we show that FSS gives estimates for $\nu$ and $W_c$ in agreement with the results from TMM.

\section{Results}\label{sec:res}

In this chapter, we focus mainly on the first two representative cases of the lattice family, $\mathcal{L}_3(1)$ and $\mathcal{L}_3(2)$. Note that, due to approximate mirror symmetry of the energy spectrum around $E=0$ (which is exact when $\varepsilon_{\vec{X}}^{(L)}=0$), we show results only for positive energies $E\geq 0$, although we have computed data for the full spectrum. 

\subsection{The Lieb lattice $\mathcal{L}_3(1)$}\label{sec:L31}

In this first case, a single macroscopic degeneracy of CLS exists at $E=0$. 
We therefore begin to study the localization lengths $\Lambda_M$ via TMM at energy $E \neq 0$, in order to avoid possible complications of the numerical schemes due to the degeneracy. 

\subsubsection{Existence of localization transitions}
\label{sec:loctrans}

\begin{figure*}[tb]
    \centering
    (a)\includegraphics[width=0.63\columnwidth]{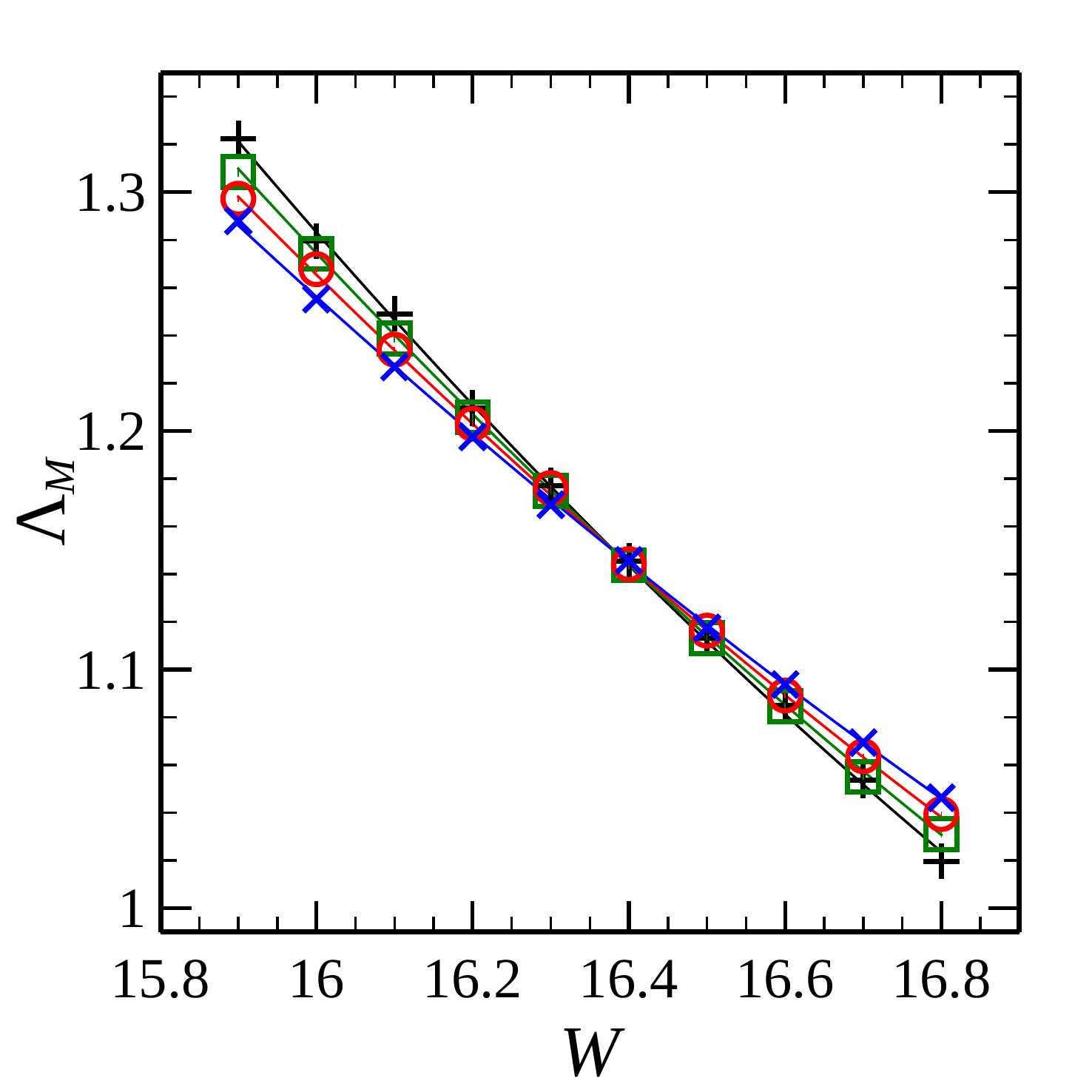}
    (b)\includegraphics[width=0.63\columnwidth]{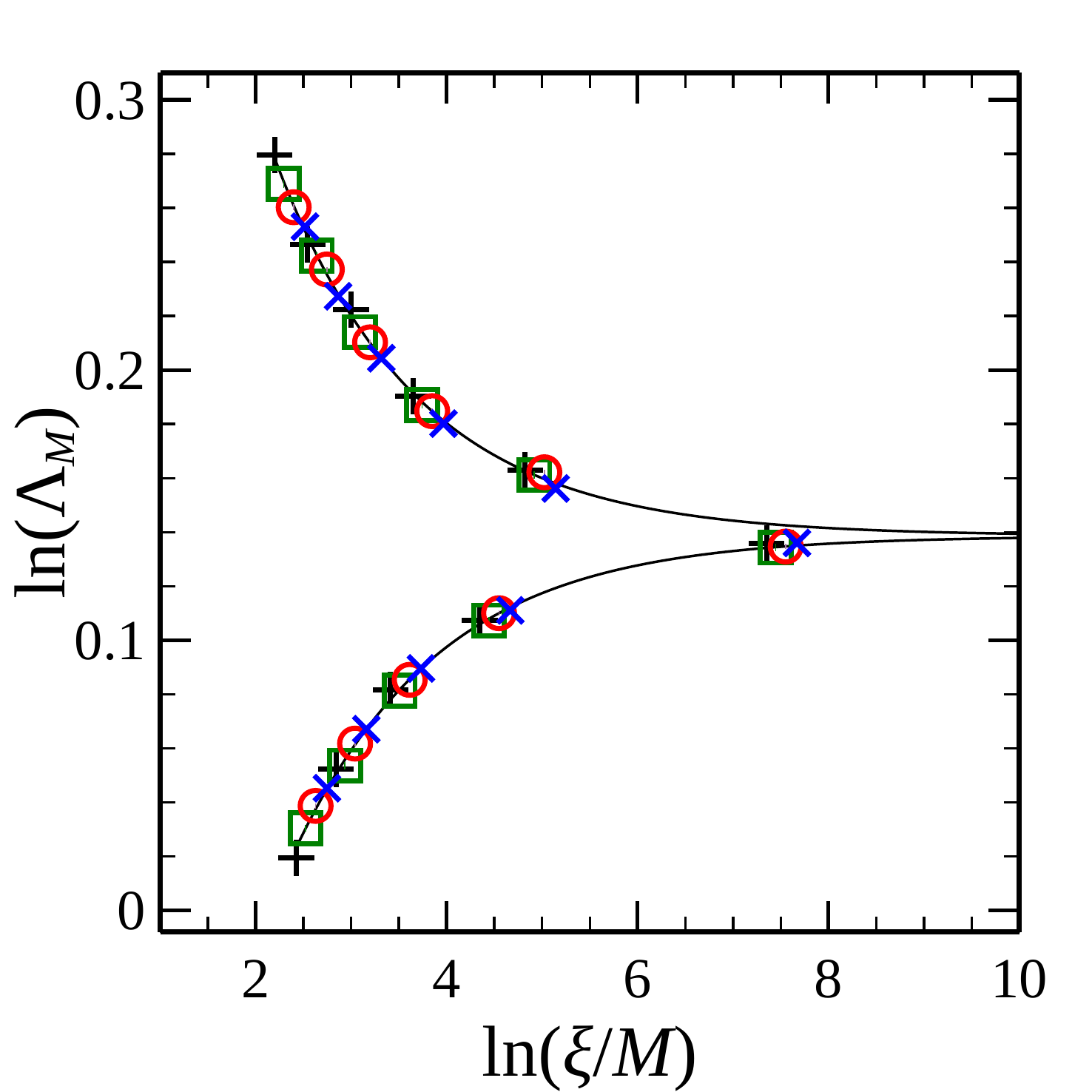}
    (c)\includegraphics[width=0.63\columnwidth]{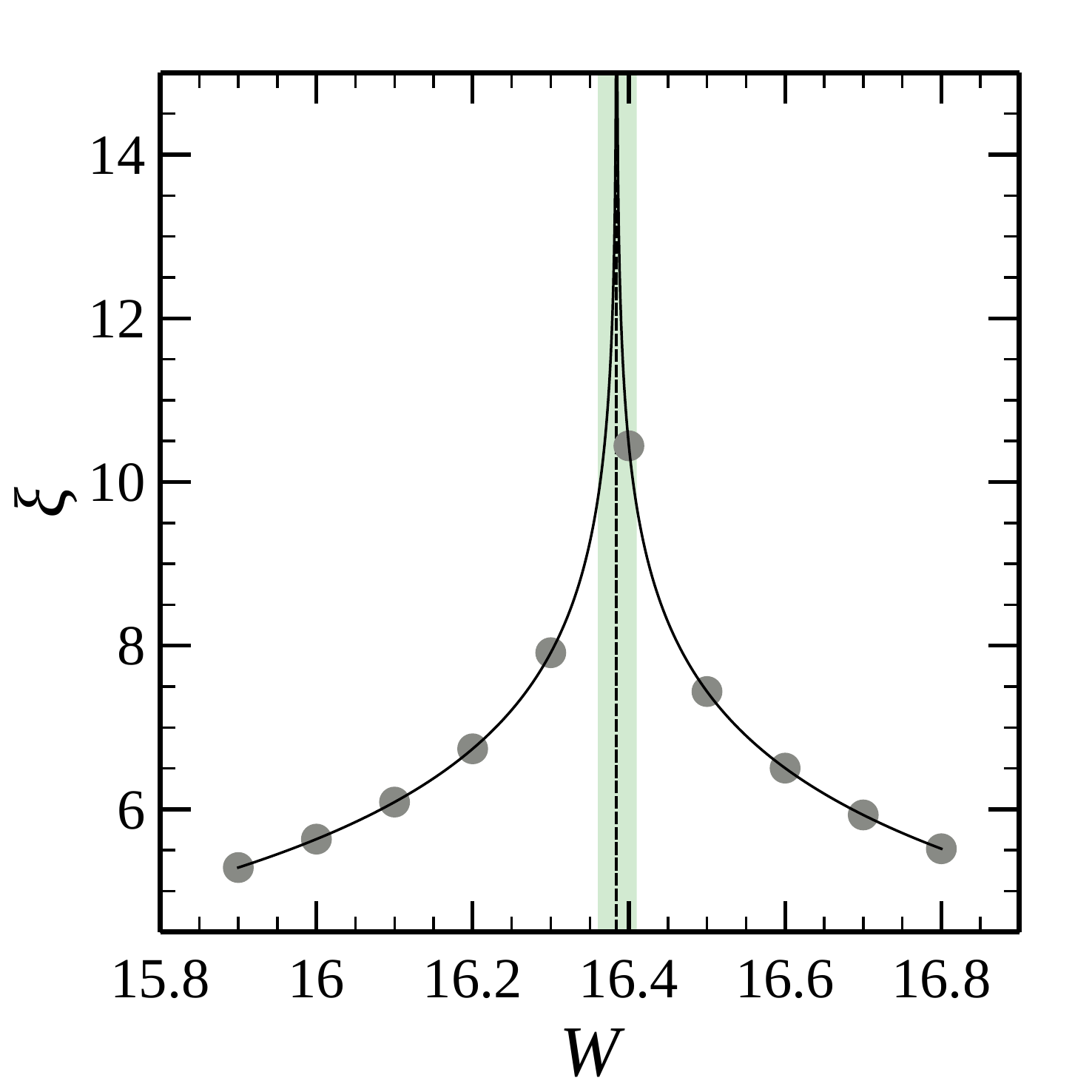}
    (d)\includegraphics[width=0.63\columnwidth]{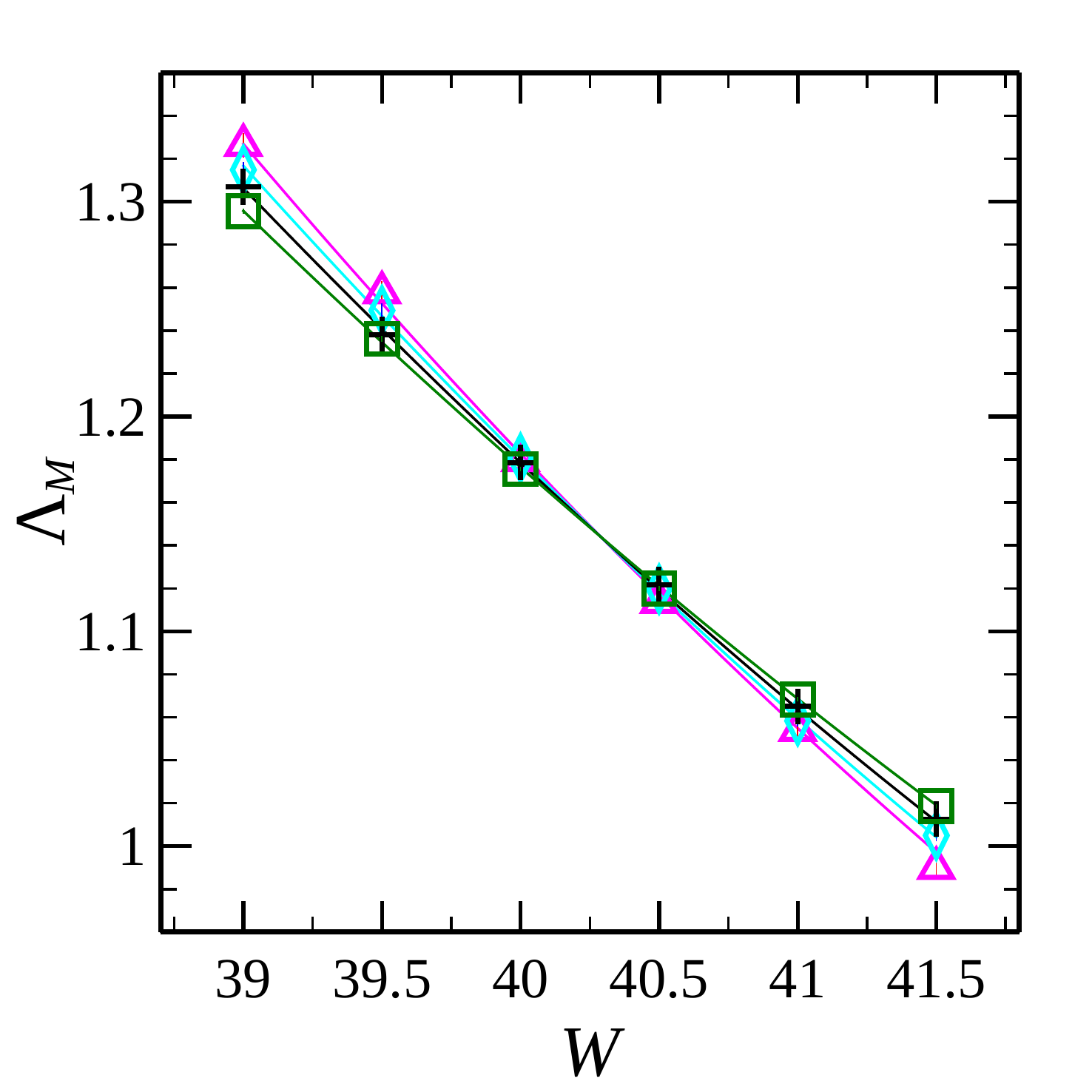}
    (e)\includegraphics[width=0.63\columnwidth]{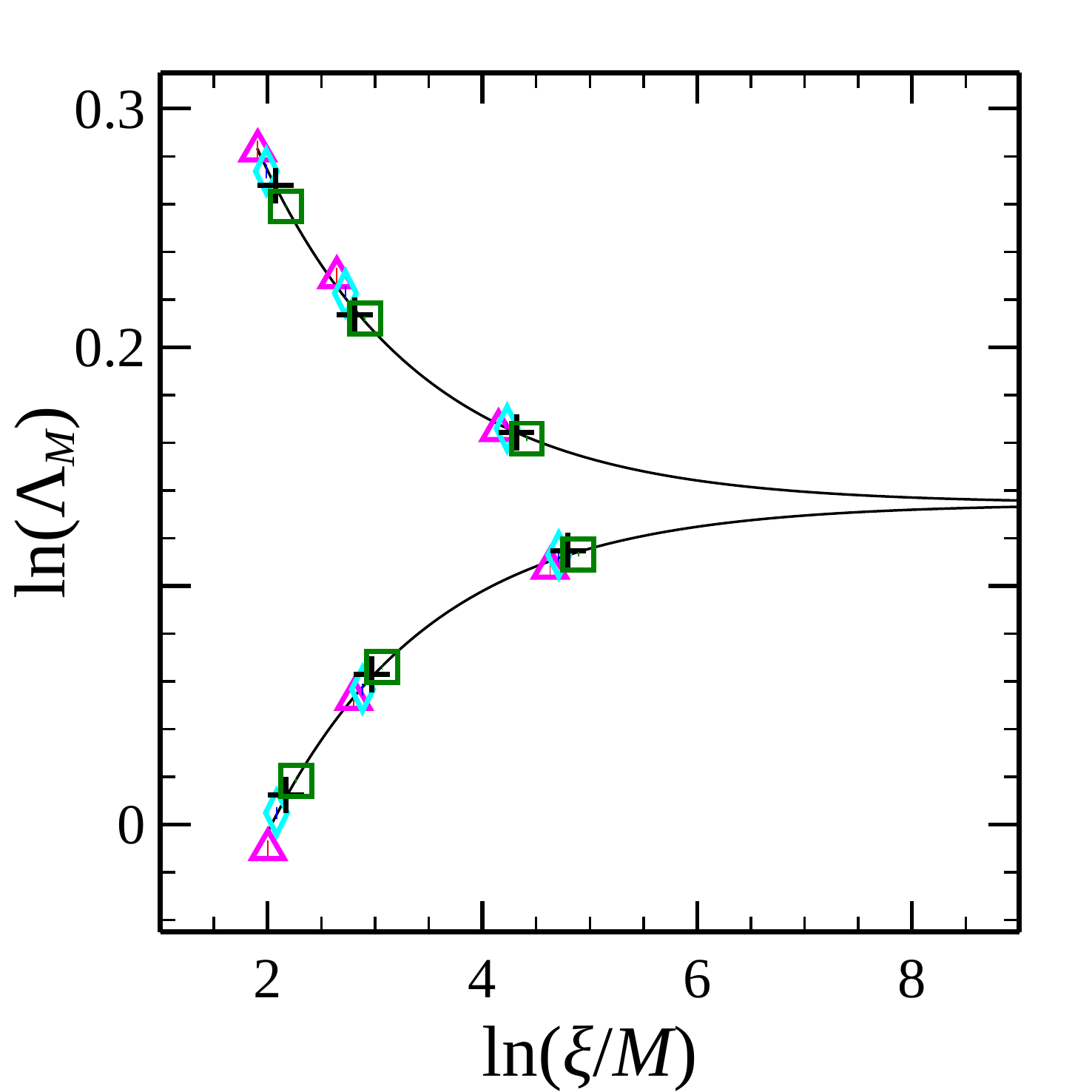}
    (f)\includegraphics[width=0.63\columnwidth]{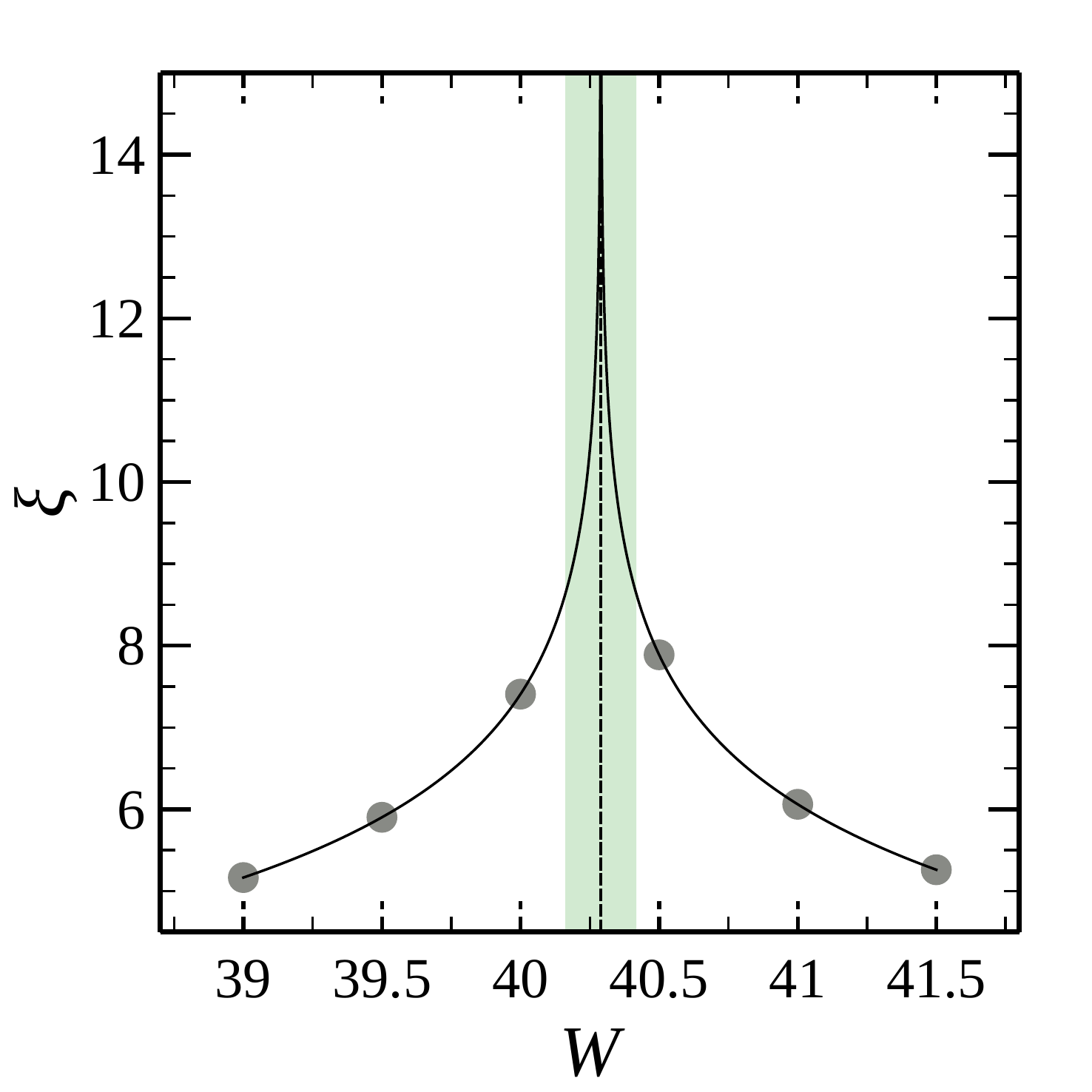}
    \caption{Finite-size scaling of the reduced localization lengths $\Lambda_M$ for $\mathcal{L}_3(1)$ at large $W$ regimes with $E=1$ (a,b,c) and $E=0.4$ (d,e,f), respectively. The bar area $M^2$ ranges from $16^2$ (blue $\times$), $18^2$ (red $\bigcirc$) with maximal convergence error $\leq 0.1\% $, to $20^2$ (green $\Box$), $22^2$ (black $+$) with maximal convergence error $\leq 0.22\%$, and to $24^2$ (cyan $\Diamond$) and $26^2$ (magenta $\bigtriangleup$) with maximal convergence error $\leq 0.5\%$. The reduced localization length $\Lambda_M$ versus the disorder strength $W$ on the cube sites, and the fits to the data showed in solid line with expansion coefficients $n_r=2$ and $m_r=1$ for both graphs are present in (a) and (d). The (b) and (e) give double logarithmic plot of scaling function $\Lambda_M$ and $\xi/M$ with scaled data points. The scaling parameter $\xi$ as a function of cube disorder $W$ and the scaled data points are \revision{shown in (c) and (f), with the vertical lines indicating the estimated $W_c$ values and their CI intervals in (green) shade.} Error bars are within the symbol size. Details of the scaling results are \revision{given} in Table~\ref{table:FSS_Table}.} 
    \label{fig:FSS_E1.0_E0.4}
\end{figure*}

\begin{figure*}[tb]
    \centering
    (a)\includegraphics[width=0.99\columnwidth]{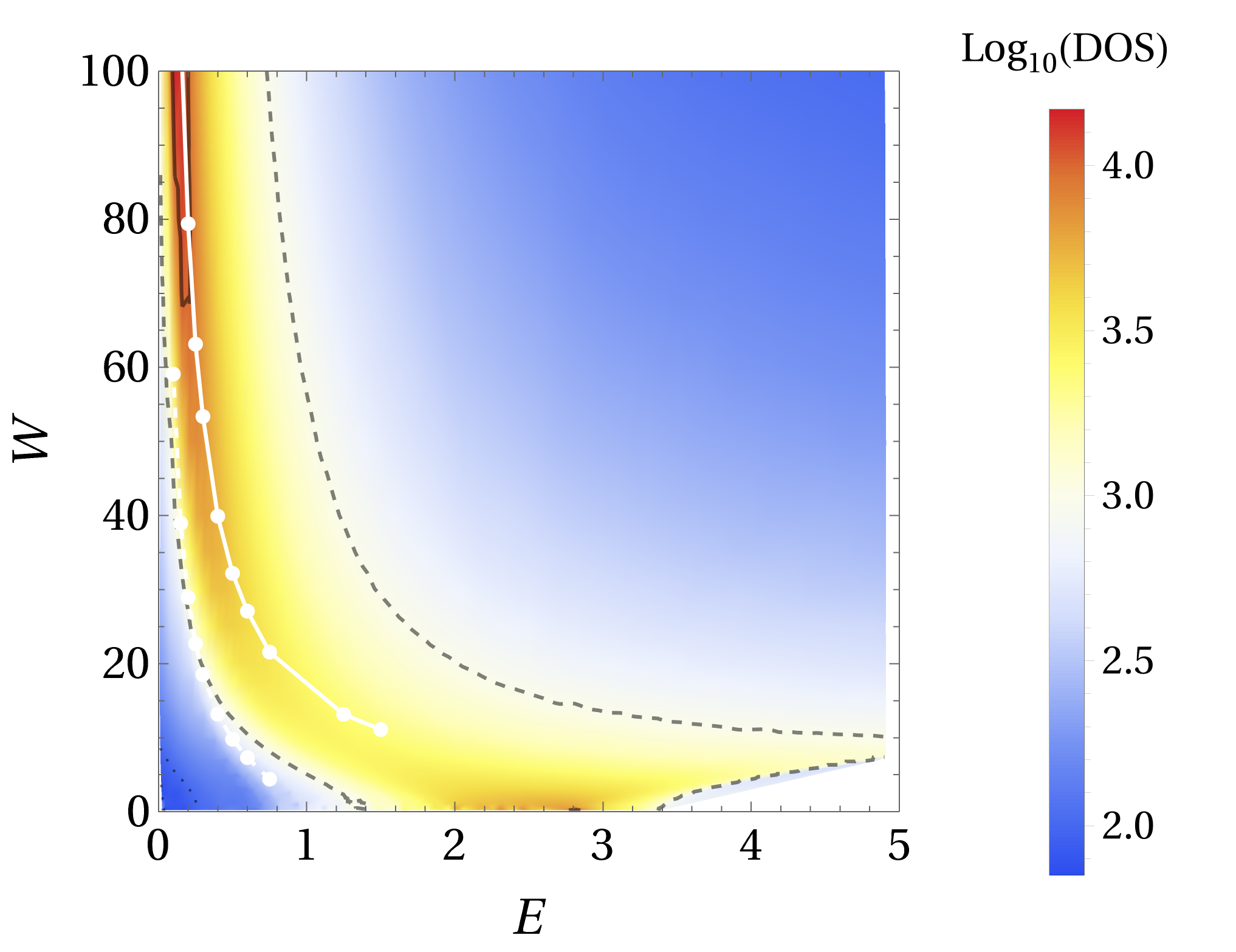}
    (b)\includegraphics[width=0.88\columnwidth]{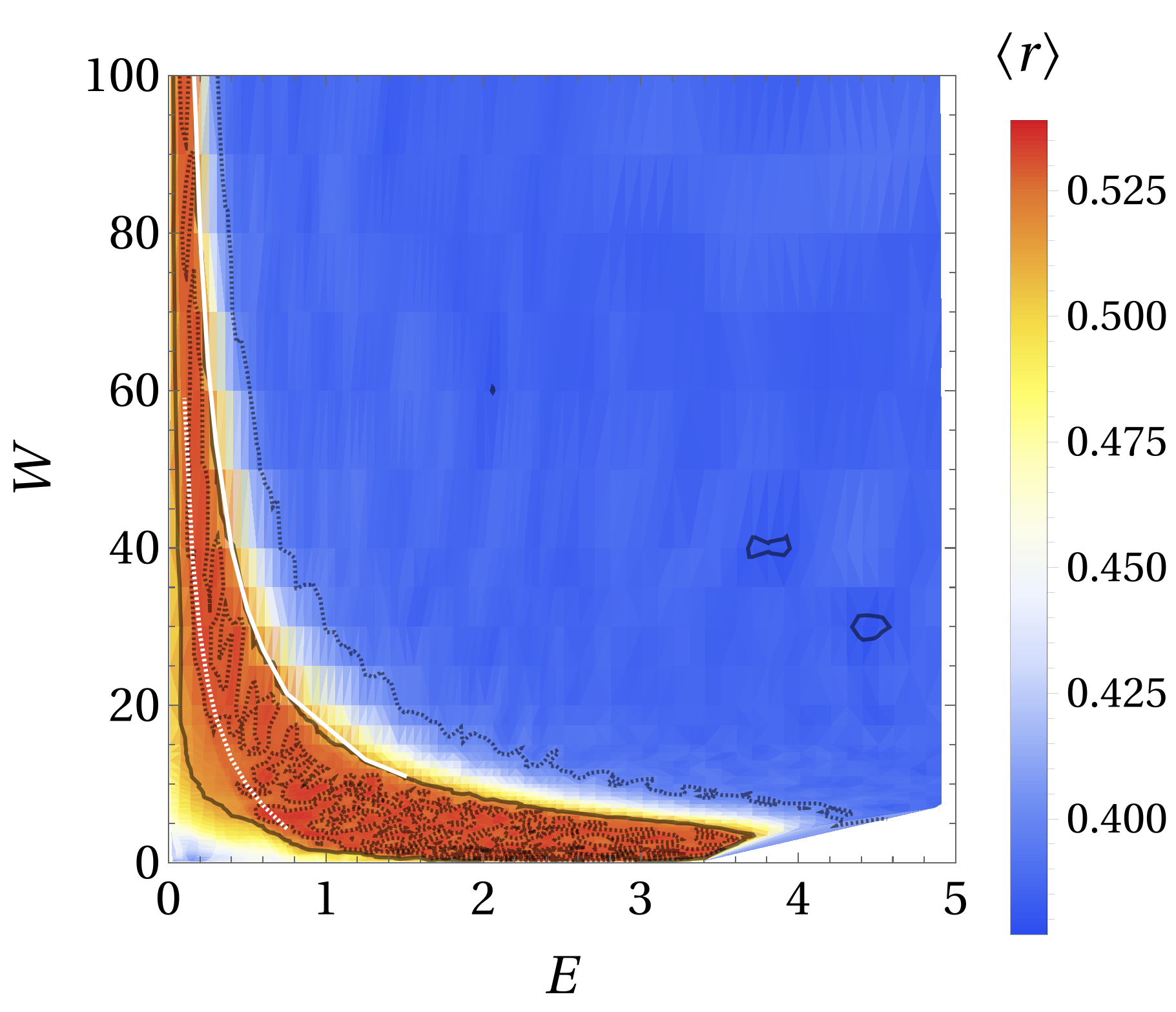}
%
    \caption{Energy $E$ and disorder $W$ dependant (a) DOS and (b) $r$-values for $\mathcal{L}_{3}(1)$. The diagrams have been obtained for system sizes $L=4\times 20^3$ and at least $100$ independent potential configurations for each $(E,W)$ pair. For each such configuration, up to $100$ energy eigenvalues around the target energy $E$ have been computed. The minimal energy spacing is $\Delta E=0.05$ while the minimal disorder spacing $\Delta W=0.1$; adaptively chosen $3300$ individual $(E,W)$ pairs contribute to the computed DOS and $r$-value density plots.
    The flat-band states at $E=0$ ($\leq 10^{-10}$) are not shown in both panels for clarity.
    The dark lines for (a) denote contours of $10^3$ (dashed) and $10^4$ (solid) states, while for (b) the lines in the red-shaded region correspond to $\langle r\rangle=0.53$ (dashed),  $0.5145$ (solid) and in the blue-shaded region, they denote $\langle r\rangle=0.4$ (dashed) and $0.38$ (solid).
    The white lines in (a) and (b) denote estimates of the transitions obtained by small-$M$ TMM with the $2$ different lines corresponding to the crossings of $\Lambda_M$ values between $M=6$ and $M=8$ 
    from localized-to-delocalized (solid) and delocalized-to-localized (dashed) 
    behaviour upon decreasing $W$ at constant $E$. In (a) these small-$M$ estimates for $(E_c,W_c)$ are given as white circles.
    }
    \label{fig:dos_r_l31}
\end{figure*}

In Fig.~\ref{fig:FSS_E1.0_E0.4}(a) we show the localization length $\Lambda_M$ at energy $E=1$ for $M^2$ ranging from $16^2$ to $22^2$ computed with high precision. 
These curves show a stable intersection point, indicating the existence of a critical disorder $W_c$ separating metallic from localized phases. Such critical transition is extracted by \revision{FSS} shown in Fig.~\ref{fig:FSS_E1.0_E0.4}(b,c), yielding $W_c=16.38(2)$ -- which incidentally is roughly the same as the standard Anderson transition $W_c=16.590(12)$ for a cubic lattice at $E=0$~\cite{Rodriguez2011MultifractalTransition}. 
However, the same computation repeated closer to the macroscopic degeneracy -- namely at $E=0.4$, as shown in Fig.~\ref{fig:FSS_E1.0_E0.4}(d-f)  -- yields a substantially higher critical transition value $W_c\approx 40.29(7)$ than the value for $E=1$ \cite{Note5}.
%
These two results seem to hint towards a divergence of $W_c$ as the energy approaches the macroscopic degeneracy at $E=0$. Consequently, we  systematically estimate the critical transition $W_c(E)$ within the interval $0< E \leq 1.5$ -- \emph{i.e.}\ strictly different than $E = 0$ -- for small system sizes $M=6,8$ via TMM with maximal convergence error of $\leq 0.5\%$. 
The resulting curve is shown in Fig.~\ref{fig:dos_r_l31}(a) with white circles connected by a solid line within the yellow region -- confirming the divergence of $W_c(E\rightarrow 0)$. 

\subsubsection{Spectral characterization of the localization transitions}

To further validate the behaviour of $W_c(E)$ and to compute the overall phase-diagram, we look at the spectral properties of the Hamiltonian \eqref{Equ:def1}. Details on the computations are reported in the caption of Fig.~\ref{fig:dos_r_l31}. 
The DOS -- shown in Fig.~\ref{fig:dos_r_l31}(a) for $E>0$ and different disorder strengths $W$ -- exhibits intriguing phenomena close to the macroscopic degeneracy level $E=0$ in both the weak and the large $W$ regimes. Namely, we observe 
(i) a depletion of the DOS at $E\leq 1$ for $W\rightarrow 0$, and 
(ii) a strong enhancement of the DOS at $E\leq 1$ for $W\rightarrow \infty$. 

\begin{figure}[bt]
    \centering
    \includegraphics[width=0.95\columnwidth]{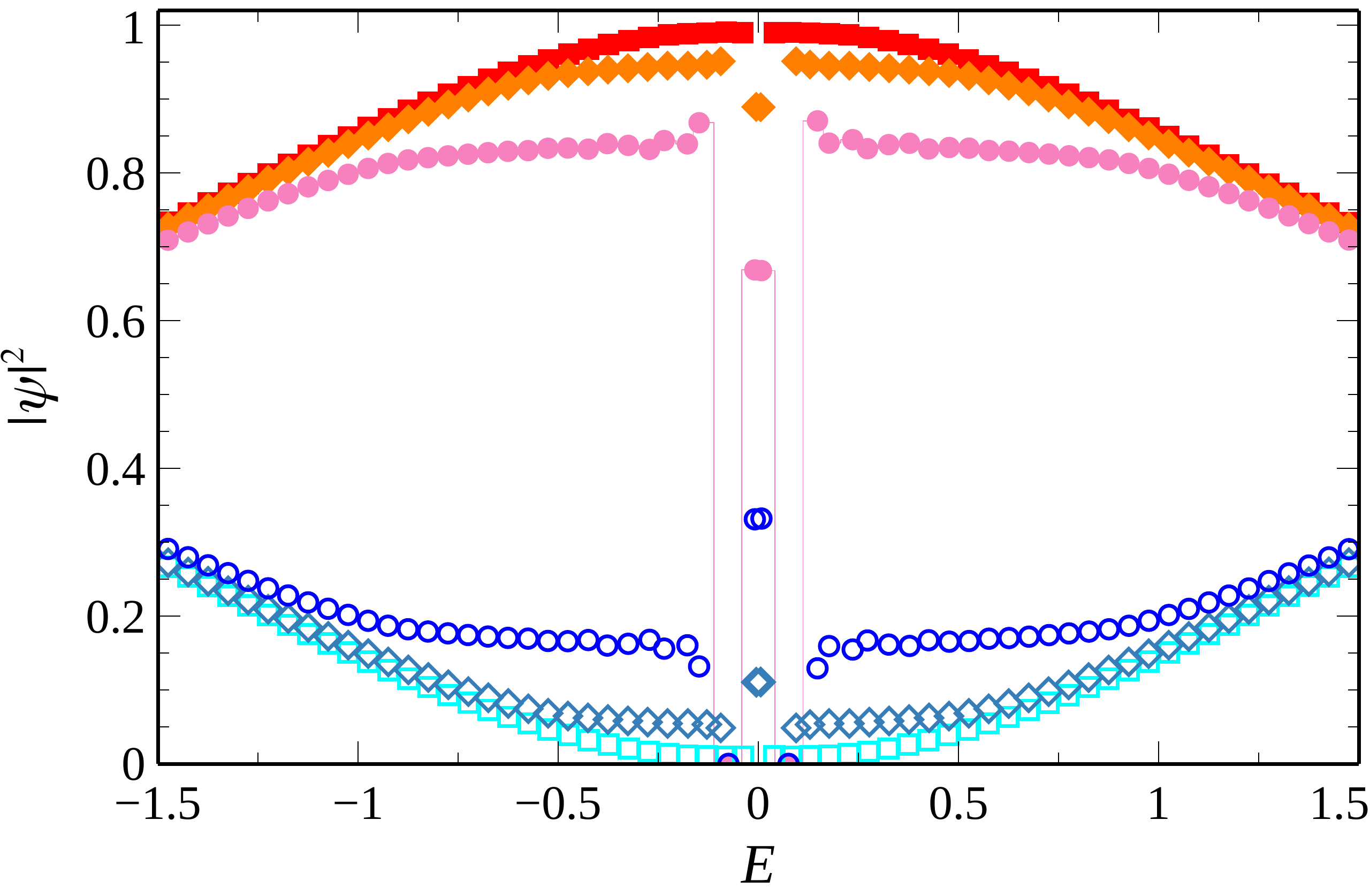}
    \caption{Projected probabilities $|\psi(x)|^2$ for cube sites (blue colors, open symbols) and Lieb sites (red colors, filled symbols) with disorders $W=10$ ($\circ$), $20$ ($\Diamond$) and $50$ ($\Box$) for ${\cal L}_3(1)$. 
    The line for Lieb sites with $W=10$ is given to highlight that the data points represent averages for $144$ potential configurations with energy resolution $\Delta E = 0.05$. The system size in all cases is $L= 4 \times 20^3$.}
    \label{fig:probs_l31_2}
\end{figure}

The former observation (i) is related to the fact that in the clean case $W=0$ the flatband $E=0$ is touching the remaining dispersive bands via conical intersections -- with consequent decrease of the DOS as $E\rightarrow 0$, as discussed in Ref.~\cite{Liu2020a}. 
The latter observation (ii) instead follows from the fact that for large $W$ it becomes energetically favorable for eigenstates to populate the unperturbed Lieb sites (where the CLS live) rather than the disordered cube sites. 
This is confirmed in Fig.\ \ref{fig:probs_l31_2} where we show the projected norm of eigenstates at the Lieb sites (red colors) and the cube sites (blue colors) as function of the energy $E$ for different disorder strength $W$ (shown with different symbols). 
What appears is an increase of the relative norm in the Lieb sites (and complementary a decrease of the relative norm in the cube-sites) as $|E|\rightarrow 0$ -- trends which are enhanced as the disorder $W$ increases. In particular, for strong disorder $W=50$, the norm of the eigenstates for $E\ll1$ is almost exclusively located at the Lieb sites.  
Such effective projection of the eigenstates at $E\neq 0$ on the set of CLS at $E=0$ results in lowering the energies of a large fraction of states close to the macroscopic degeneracy -- and, consequently, the strong enhancement of the DOS for $E\ll1$ as $W\rightarrow \infty$. 
Note that in these calculations we have excluded those eigenstates with $|E|\leq 10^{-4}$, removing the degenerate CLS. However, close to $E=0$, each of the different potential realizations yields a single eigenstate at $E\sim 10^{-2}$, which is an accidental degeneracy following from the  $\langle \varepsilon_{\vec{X}}^{(c)} \rangle_{N\rightarrow \infty} \rightarrow 0$ \cite{Ramachandran2017}. 
Such eigenstates result in the outlier points close to $E=0$ in Fig.\ref{fig:probs_l31_2}. 

\subsubsection{Spectral gap ratio statistics}

%

The diverging behaviour of $W_c(E)$ shown in Fig.~\ref{fig:dos_r_l31}(a) had been estimated via TMM. In order to find further support for this behaviour, we now use the independent spectral gap ratio statistics for $\langle r \rangle$ outlined in section \ref{sec:model} to compute the full phase-diagram for $\mathcal{L}_3(1)$ via sparse-matrix diagonalization. 
In Fig.~\ref{fig:dos_r_l31}(b) we show the $\langle r \rangle$ for $L = 4\times 20^3$ as function of the $E$ and $W$. 
The results convincingly confirm the diverging trend for the transition curve $W_c(E)$ from extended with $\langle r \rangle\sim \langle r \rangle_\text{Sur}$ ($\sim 0.53$) to localized with $\langle r \rangle \sim \langle r \rangle_\text{Poi}$ ($\sim 0.38$) as $E\rightarrow 0$ for $W \gtrsim 10$. In particular,
the $r$-value-based transition line shows strong agreement with the transition curve obtained from the scaling behaviour of localization lengths $\Lambda_M$ (shown in Fig.~\ref{fig:dos_r_l31}(b) with white solid line). Furthermore, we observe that close to $E=0$ in the small $W$ regime, the $\langle r\rangle$ drops from $r\sim 0.529$ -- a decrease occurring in correspondence to the depletion of the DOS. 

\begin{figure*}[tbh]
    \centering
    (a)\includegraphics[width=0.63\columnwidth]{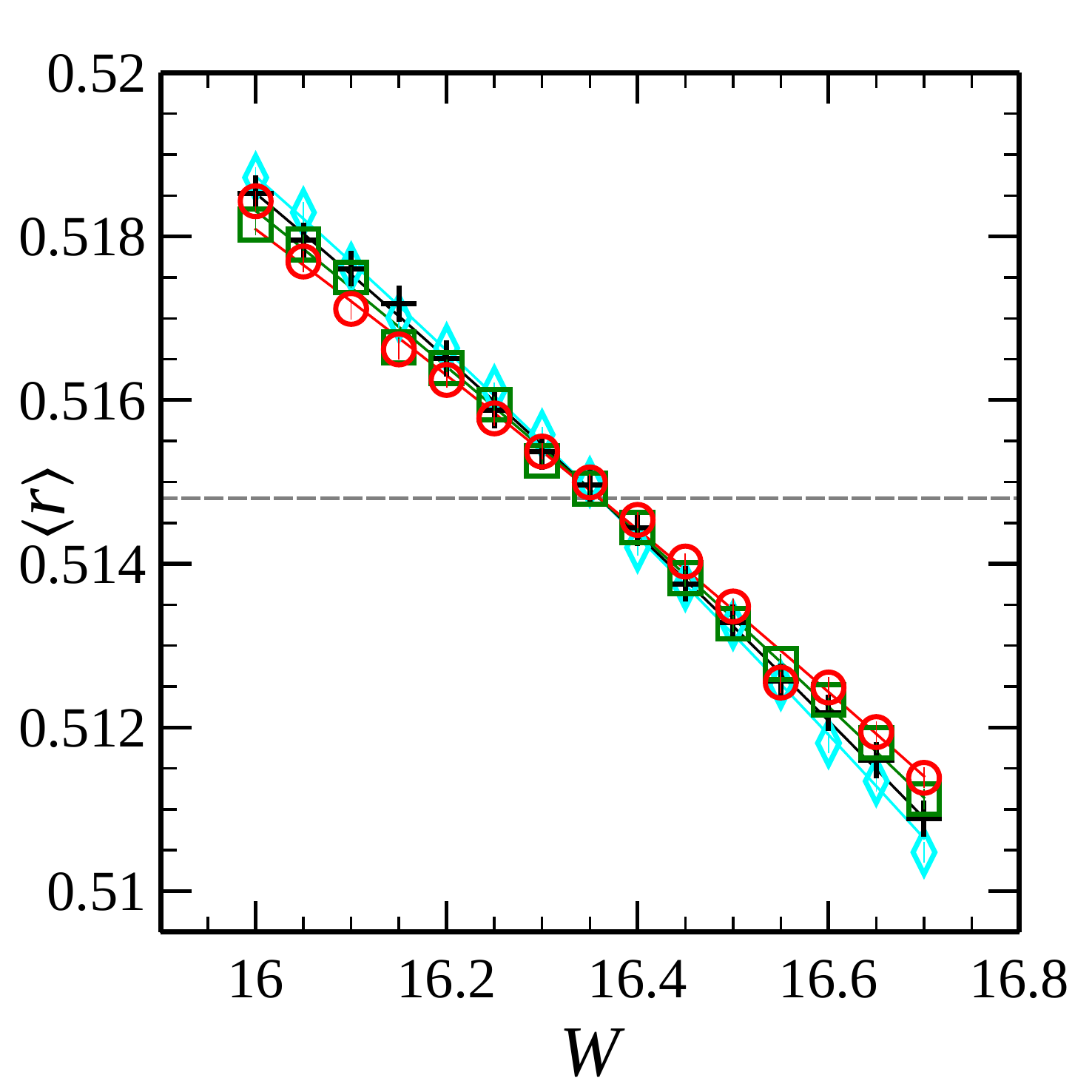}
    (b)\includegraphics[width=0.63\columnwidth]{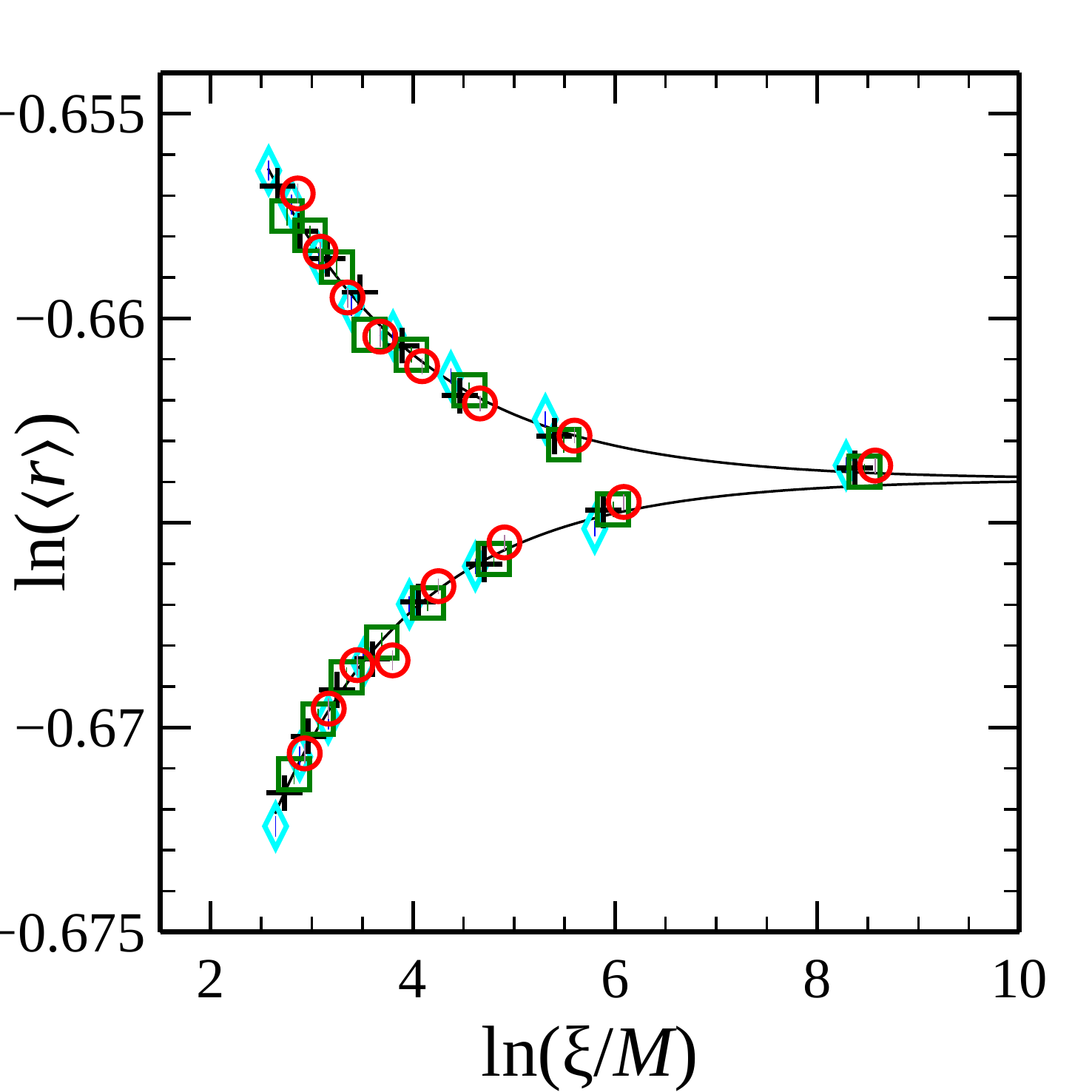}
    (c)\includegraphics[width=0.63\columnwidth]{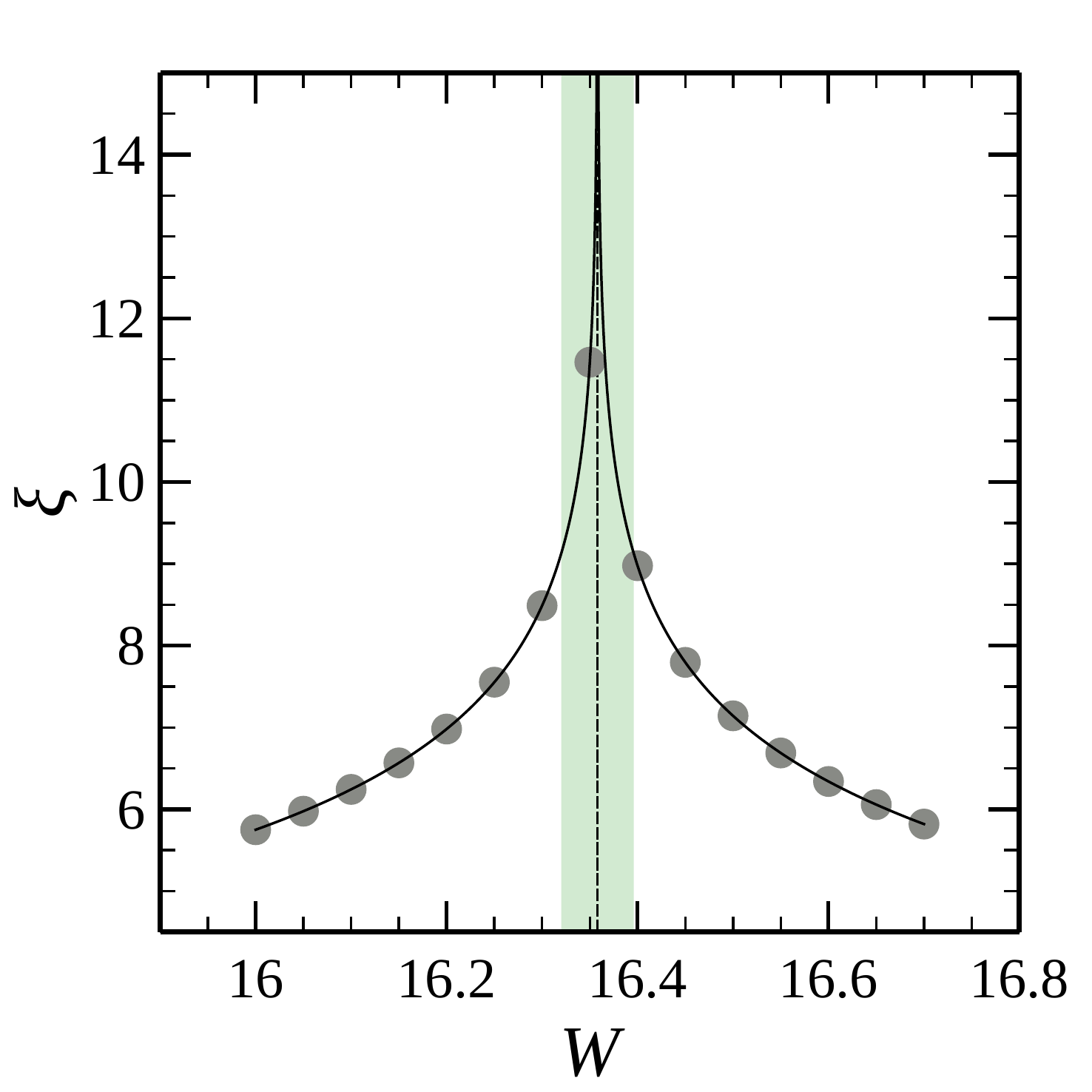}
    (d)\includegraphics[width=0.63\columnwidth]{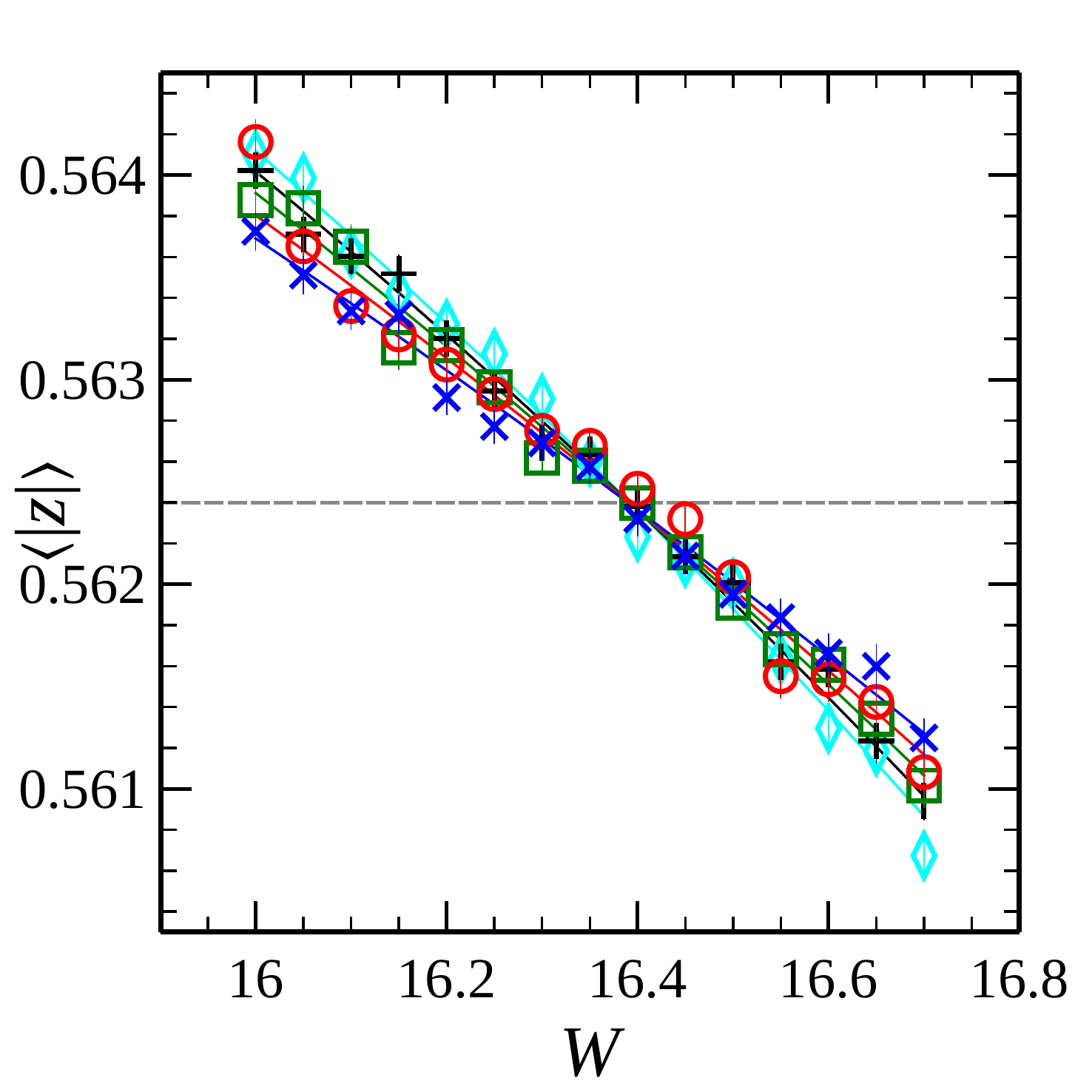}
    (e)\includegraphics[width=0.63\columnwidth]{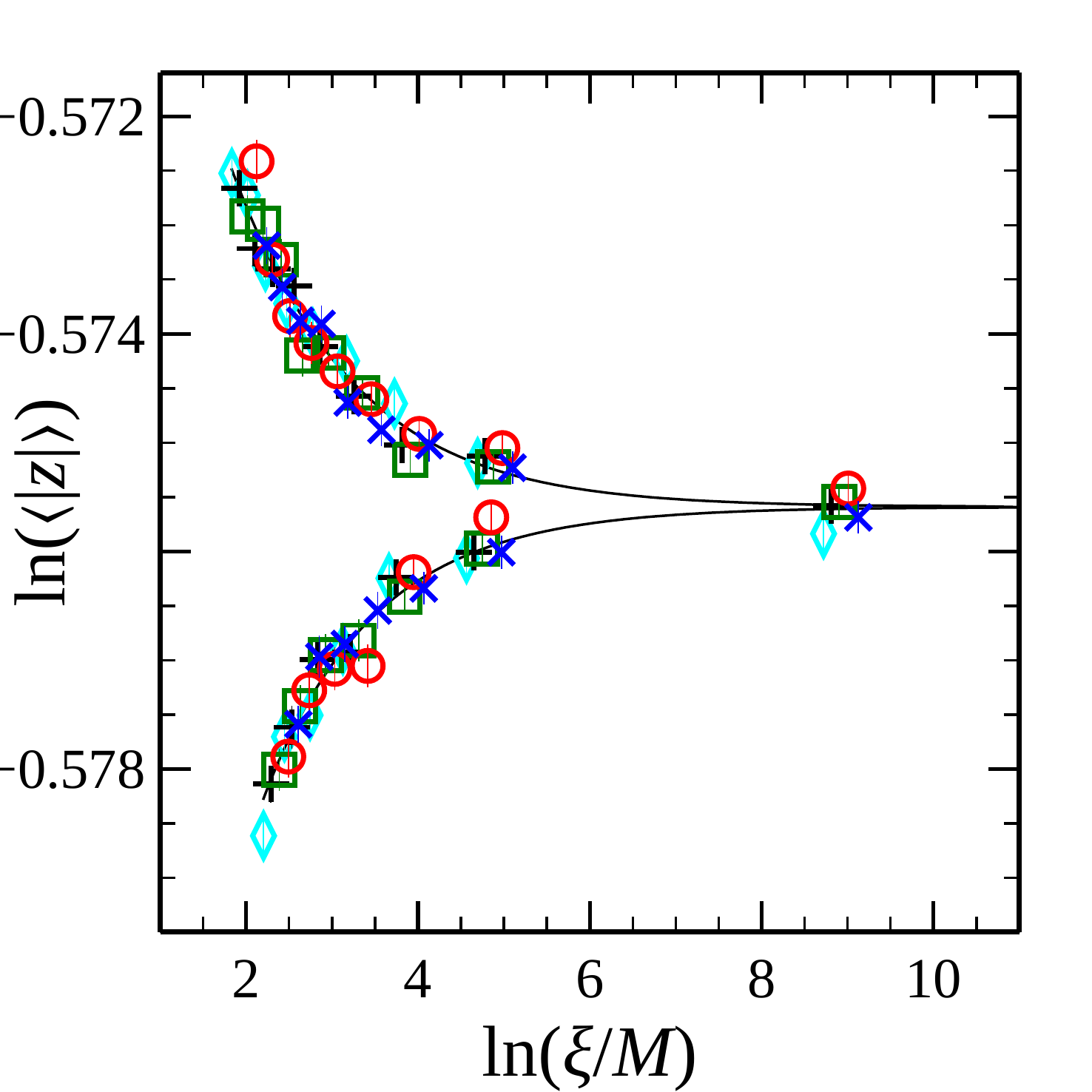}
    (f)\includegraphics[width=0.63\columnwidth]{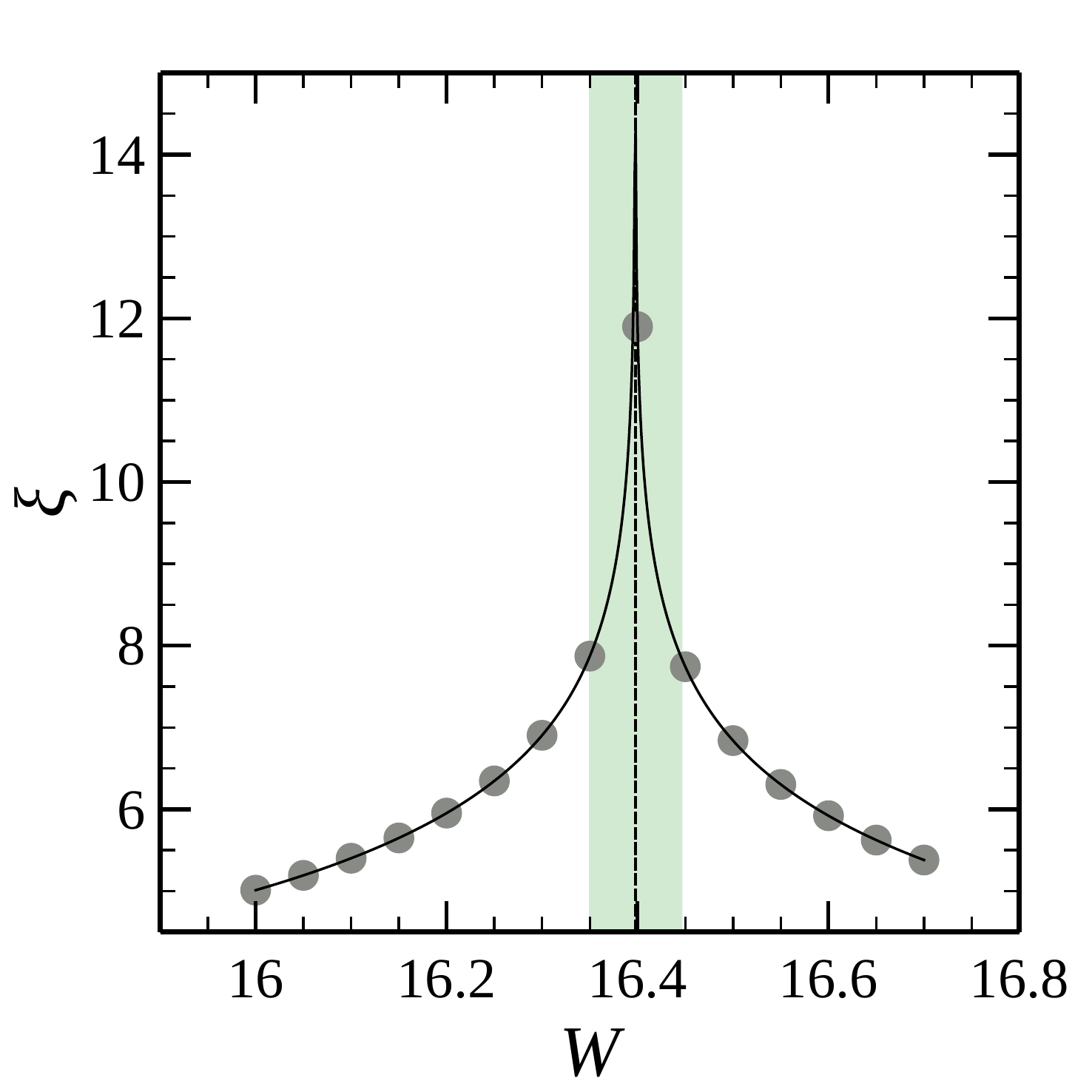}
\caption{Finite size scaling results of the (a--c) $r$-values and (d--f) $z$-values for $\mathcal{L}_3(1)$ in the high $W$ region at $E=1$, respectively.
    The system sizes $L= 4 \times N^3$ are from $N=$ %
    $16$ (blue $\times$), $18$ (red $\bigcirc$), $20$ (geen $\Box$), $22$ (black $+$), and to $24$ (cyan $\Diamond$). %
    For each $N$ and $W$, $10000$ different potential configurations have been calculated and for each we include up to $100$ energy eigenvalues around the target energy $E$ in the computing $\langle r \rangle$ and $\langle |z| \rangle$.  %
    Panel (a) and (d) show the $\langle r \rangle$-values ($\langle |z|\rangle$-values) versus $W$ data and the fits to the data, as given in Table \ref{tab:L31_FSS} with expansion coefficients $n_r=2$ and $m_r=1$, are marked with solid lines. Panels (b) and (e) give double logarithmic plots of scaling function $r$-values($z$-values) versus $\xi/M$ with scaled data points. The scaling parameter $\xi(W)$  is shown in panels (c) and (f). \revision{The vertical lines indicate the $W_c$ values and the shaded region their CI.} 
    Error bars as shown in panels (a), (b), (d) and (e) are mostly within symbol size.
    The horizontal lines in panels (a) and (d) denote the values $\langle r \rangle= 0.5148$ (with CI $[0.5143,0.5153]$)
    and $\langle |z|\rangle=0.5624$ (with CI $[0.5621,0.5627]$)
    obtained at $W_c$.} 
     \label{fig:rz_vs_W_l31_E0100}
\end{figure*}

We first have a more in-depth look at the localization-to-delocalization transition (white solid line in Fig.~\ref{fig:dos_r_l31}(b)) when starting from high $E$ and/or $W$ values. Analogously with the TMM, we fix energy to $E=1$ and study the behaviour of $\langle r\rangle(W)$ for various system sizes $N$. 
In Fig.~\ref{fig:rz_vs_W_l31_E0100}(a-c) we show FSS results for $\langle r\rangle$ values for $N$ ranging from $18$ to $24$ around the expected transition value $W_c\approx 16.4$. 
Note that the errors bars (of order $10^{-3}$) are obtained as standard error $\sqrt{([\langle r\rangle^2]-[\langle r\rangle]^2)/(R-1)}$, where $[\ ]$ denotes the average over $R$ potential realization, and $\langle \rangle$ represents the average within a given potential realization~\cite{Oganesyan2007c}. 
We find that the critical disorder $W_c\approx 16.36(2)$ as computed with the $r$-statistics is in excellent agreement with the critical value $W_c=16.38(2)$ obtained via TMM. Furthermore, as detailed in Table~\ref{table:FSS_Table}, the FSS results are in agreement with the conventional critical exponents for the 3D Anderson transition \cite{Slevin1999b,Rodriguez2011MultifractalTransition}. 
Indeed, the critical exponent $\nu=1.51$ is also in agreement with the TMM result. 
Last, we see in Fig.\ \ref{fig:rz_vs_W_l31_E0100}(a) that the value $\langle r \rangle (W_c)= 0.5145$, which is one of the contour lines highlighted in Fig.\ \ref{fig:dos_r_l31}(b), separates localized from extended behaviour, again emphasizing the consistency of our results.

\begin{table*}
\centering
\setlength{\tabcolsep}{3.25mm}%
\begin{tabular}{cccccccccc}
\hline \hline \noalign{\smallskip}
\multicolumn{9}{c}{ Reduced localization length $\Lambda_M$}  \\
    $M$&    $E$&   $\delta W$&     $n_{r}$ $m_{r}$ &
    $W_{c}$&  CI$(W_{c})$&  $\nu$&  CI$(\nu)$&  $p$        \\ 
    20,22,24,26&    0.4&    39.0-41.5&       $2\ \ $ $1\ \ $&  $\textbf{\emph{40.29}}$&   $[40.16,40.42]$& $\textbf{\emph{1.50}}$&  $[1.28,1.73]$&  $0.52$ \\ 
    20,22,24,26&    0.4&    39.0-41.5&       $3\ \ $ $1\ \ $&  $40.29$&   $[40.15,40.43]$&   $1.51$&  $[1.27,1.75]$&  $0.46$   \\
    20,22,24,26&    0.4&    39.0-41.5&       $3\ \ $ $2\ \ $&  $40.35$&   $[40.14,40.56]$&   $1.51$&  $[1.26,1.76]$&  $0.44$   \\
    20,22,24,26&    0.4&    39.0-41.5&       $4\ \ $ $1\ \ $&  $40.30$&   $[40.16,40.44]$&   $1.51$&  $[1.27,1.75]$&  $0.43$   \\
    Averages:   & & & &     $40.31(4)$&   &                                                  $1.51(6)$&    \\[2ex]

    $M$&    $E$&   $\delta W$&     $n_{r}$ $m_{r}$&
    $W_{c}$&    CI$(W_{c})$&    $\nu$&    CI$(\nu)$&    $p$        \\ 
    16,18,20,22&    1&    15.9-16.8&       $2\ \ $ $1\ \ $&   $\textbf{\emph{16.38}}$&   $[16.36,16.41]$&    $\textbf{\emph{1.50}}$&  $[1.37,1.63]$&  $0.22$ \\ 
    16,18,20,22&    1&    15.9-16.8&       $3\ \ $ $1\ \ $&   $16.39$&   $[16.36,16.41]$&    $1.51$&     $[1.38,1.65]$&   $0.19$      \\
    16,18,20,22&    1&    15.9-16.8&       $3\ \ $ $2\ \ $&   $16.41$&   $[16.38,16.45]$&    $1.50$&     $[1.37,1.63]$&   $0.37$      \\
    16,18,20,22&    1&    15.9-16.8&       $4\ \ $ $1\ \ $&   $16.39$&   $[16.36,16.42]$&    $1.51$&     $[1.37,1.65]$&   $0.18$      \\
    Averages:    & & & &                                      $16.39(1)$&   &                $1.51(4)$&    \\    

\\[0.5ex]\hline\noalign{\smallskip}
\multicolumn{8}{c}{$r$-Values }\\
    $N$&    $E$&   $\delta W$&     $n_{r}$ $m_{r}$&
    $W_{c}$&  CI$(W_{c})$&    $\nu$&   CI$(\nu)$&   $p$        \\ 
    18,20,22,24&    1&    16.0-16.7&      $2\ \ $ $1\ \ $&   $\textbf{\emph{16.36}}$&  $[16.32,16.40]$&  $\textbf{\emph{1.51}}$&  $[1.21,1.80]$&   $0.56$ \\ 
    18,20,22,24&    1&    16.0-16.7&      $3\ \ $  $1\ \ $&  $16.36$&  $[16.31,16.40]$&   $1.54$&    $[1.22,1.86]$&   $0.54$     \\
    18,20,22,24&    1&    16.0-16.7&      $3\ \ $  $2\ \ $&  $16.37$&  $[16.32,16.42]$&   $1.55$&    $[1.22,1.88]$&   $0.53$     \\
    18,20,22,24&    1&    16.0-16.7&      $4\ \ $  $1\ \ $&  $16.36$&  $[16.31,16.40]$&   $1.54$&    $[1.22,1.86]$&   $0.51$      \\
    Averages:      & & & &                                   $16.36(2)$&              &   $1.54(9)$& \\
\\[0.5ex]\hline\noalign{\smallskip}
\multicolumn{8}{c}{$|z|$-Values}\\
    $N$&    $E$&   $\delta W$&   $n_{r}$ $m_{r}$&
    $W_{c}$&   CI$(W_{c})$&    $\nu$&   CI$(\nu)$&   $p$ \\ 
    16,18,20,22,24&    1&    16.0-16.7&    $2\ \ $ $1\ \ $&  $\textbf{\emph{16.40}}$&   $[16.34,16.45]$&   $\textbf{\emph{1.35}}$&  $[1.01,1.68]$&  $0.67$ \\ 
    16,18,20,22,24&    1&    16.0-16.7&    $3\ \ $ $1\ \ $&  $16.40$&   $[16.34,16.45]$&   $1.49$&  $[1.10,1.88]$&  $0.75$     \\
    16,18,20,22,24&    1&    16.0-16.7&    $3\ \ $ $2\ \ $&  $16.40$&   $[16.34,16.47]$&   $1.47$&  $[1.08,1.85]$&  $0.73$ \\
    16,18,20,22,24&    1&    16.0-16.7&    $4\ \ $ $1\ \ $&  $16.40$&   $[16.35,16.46]$&   $1.46$&  $[1.09,1.84]$&  $0.75$ \\    
    Averages:      & & & &                                   $16.40(2)$&    &              $1.44(10)$& \\
    \\
 \hline\hline
\end{tabular}
\caption{Critical parameters of the traditional (standard) Anderson transition for $\mathcal{L}_3(1)$ with reduced localization length $\Lambda_M$, $r$- and $|z|$-values as indicator, respectively. The columns give the size of the system (the width $M$ of the cross section of a TMM bar and of the side length $N$ of a cube for $\Lambda_M$ and $r$- and $|z|$-values, respectively), fixed $E$, range of $W$, FSS expansion orders $n_{r}$, $m_{r}$ and the resulting critical disorders $W_c$, their 95$\%$ confidence intervals (CI), the critical exponent $\nu$, and its CI, and the goodness of fit probability $p$ in order. The averages contain the mean of the three preceding $W_c$ and $\nu$ values, with standard error of the mean in parentheses. The bold $W_c$ and $\nu$ values highlight the fits used as examples in Fig.\ \ref{fig:FSS_E1.0_E0.4} and Fig.\ \ref{fig:rz_vs_W_l31_E0100}.}
\label{table:FSS_Table}
\label{tab:L31_FSS}
\end{table*}

These measurements are further confirmed by the results obtained via the spectral statistics based on the $|z|$ measure introduced in Refs.~\cite{Sa2020,Luo2021a}. 
In Fig.~\ref{fig:rz_vs_W_l31_E0100}(d) we plot the $\langle |z|\rangle$(W) data and corresponding FSS lines for $N$ ranging from $16$ to $24$ at $E=1$, again around the expected $W_c \approx 16.4$. In Fig.~\ref{fig:rz_vs_W_l31_E0100}(e-f) we show the associated scaling function and scaling parameter. The results agree with those obtained via $r$-statistic, albeit with larger error bars, giving a critical transition at $W_c\approx 16.40(3)$. 
Full details about the finite-size scaling(FSS) and the scaling parameters of $\Lambda_M$, $r$-values and $|z|$-values are reported in Table.\ \ref{table:FSS_Table}. \revision{In particular, we note that FSS is possible even without having to take into account irrelevant corrections to scaling. We have also performed FSS with irrelevant corrections, and found fits with acceptable $\chi^2$ statistics. However, already the FSS without irrelevant corrections is stable, i.e.\ independent of the chosen disorder range, and robust, i.e.\ $W_c$ and $\nu$ values to not violate their error boundaries when increasing the expansion orders $n_r$, $m_r$. We therefore only show the results for the latter case in Table \ref{table:FSS_Table}. This is also the case for the FSS results from the TMM data of section \ref{sec:loctrans}.}

\subsubsection{The ``inverse transition" at small $E$ and $W$ values}

As briefly mentioned above when discussing Fig.\ \ref{fig:dos_r_l31}, the region of $E\lesssim 1$ and $W\lesssim 10$ for the DOS and phase diagram of ${\cal L}_3(1)$ indicates a small DOS $\sim 10^2$ as well as small $\langle r \rangle \sim 0.4$ values. These observations suggest that the regime again corresponds to localized states, and, consequently, the system might in fact exhibit an ``inverse" Anderson transition whereby upon increasing $W$ at some fixed $0< E \lesssim 1$ one can observe a transition back into the extended regime.

\begin{figure*}[tbh]
    \centering
    (a)\includegraphics[width=0.95\columnwidth]{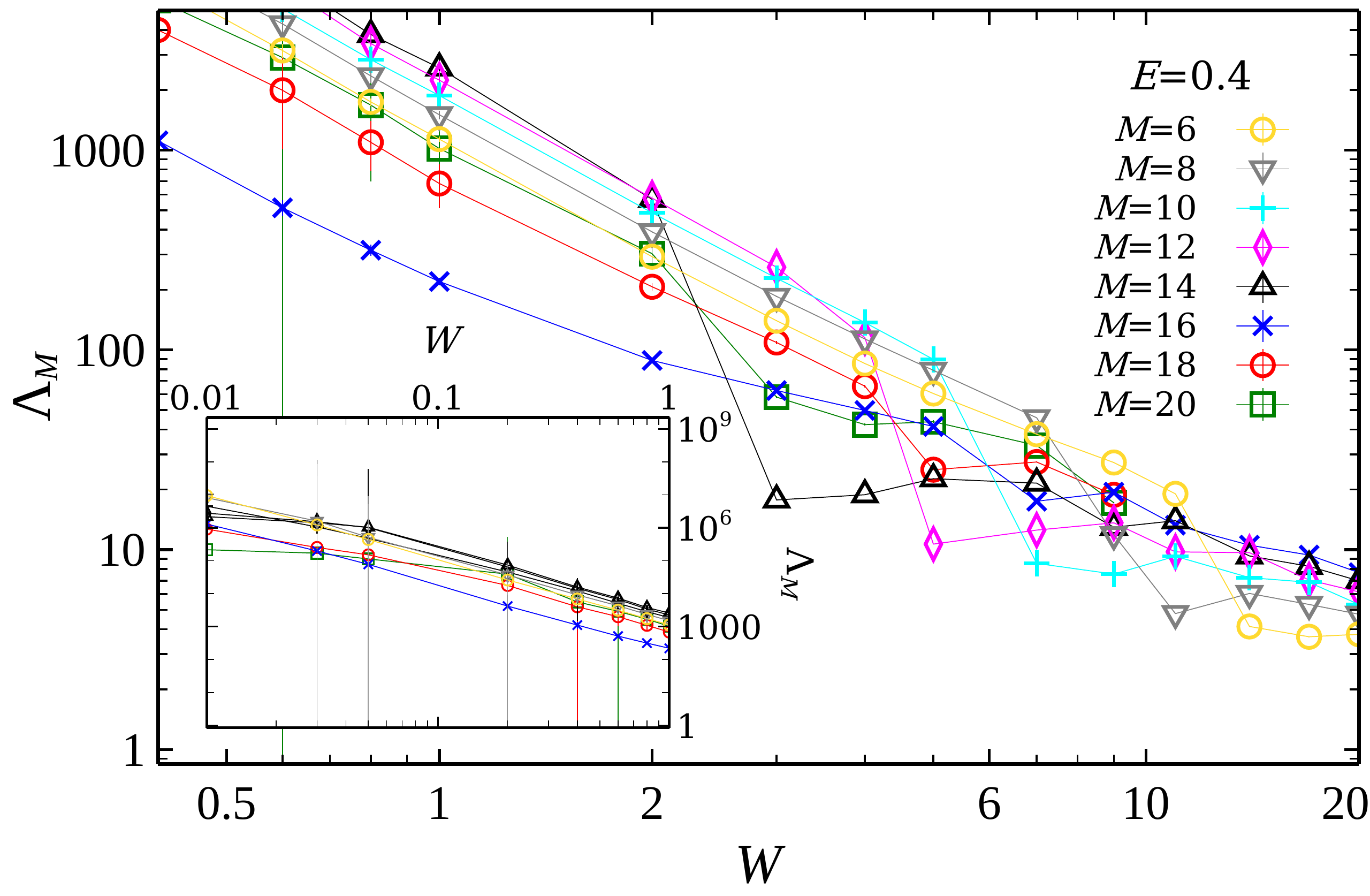}
    (b)\includegraphics[width=0.95\columnwidth]{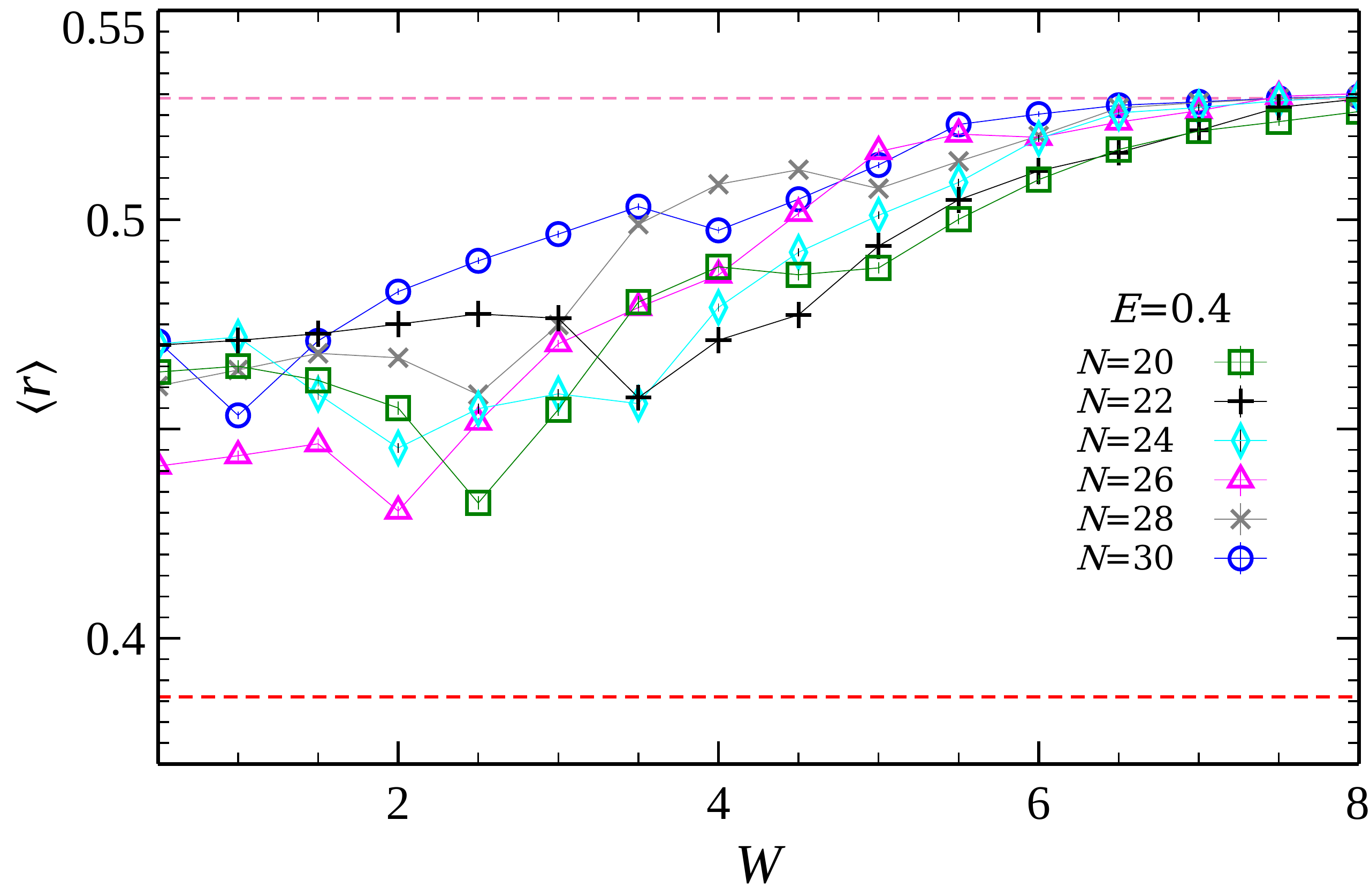}
    \caption{(a) Reduced localization length $\Lambda_M$ versus disorder $W$ at small disorder for $\mathcal{L}_3(1)$ at $E=0.4$ and TMM bar area $M=6^2$ (yellow $\bigcirc$), $8^2$ (grey $\bigtriangledown$), $10^2$ (cyan $+$), $12^2$ (magenta $\Diamond$) and $14^2$ (black $\bigtriangleup$), $16^2$ (blue $\times$), $18^2$ (red $\bigcirc$), $20^2$ (green $\Box$). 
    The error bars are all shown, and within symbol size.
    The inset focuses on the small $W$ regime $0.01\leq W\leq 1$. 
    (b) $r$-values versus disorder $W$ at $E=0.4$ for system size $L= 4 \times N^3$ with $N=20$ (green $\Box$), $22$ (black $+$), $24$ (cyan $\Diamond$) and $26$ (magenta $\bigtriangleup$), $28$ (grey $\times$), $30$ (blue $\bigcirc$) with $10000$ potential realizations for each $(N,W)$ pair. The horizontal dashed green (red) line represents the extended (localized) regime with $\langle r \rangle_\text{Sur}$ ($\langle r \rangle_\text{Poi}$). All other lines are guides to the eye, only.
    }
    \label{fig:LocLength_Inverse_E0.4_2}
\end{figure*}

In order to study this possibility in detail, we choose $E=0.4$ and again compute localization lengths $\Lambda_M$ via TMM as well as $\langle r\rangle$ statistics as function of $W$ for increasing bar width $M$ or system size $N$, respectively, aiming for a maximal convergence error of $0.1\%$ for TMM. In Fig.~\ref{fig:LocLength_Inverse_E0.4_2}, we show the resulting data. The error bars are mostly within symbol size, highlighting the reliability of the data.

We find that the localization lengths $\Lambda_M$ shown in Fig.~\ref{fig:LocLength_Inverse_E0.4_2}(a) do indeed exhibit the expected opposite dependency on $M$. For $W \leq 1$ increasing $M$ leads to an decrease of $\Lambda_M$ while for $W\gtrsim 15$, increasing $M$ increases $\Lambda_M$, at least for the larger sizes studied. Hence there seems to be indeed a change from localized behaviour at small $W$ to extended behaviour at larger $W$. 
However, we also observe considerable non-monotonic behaviour, e.g.\ for $M=14$, and a complete absence of a clearly defined crossing point to serve as estimate for $W_c$. The behaviour cannot be captured by the standard FSS techniques and the required ``corrections to scaling" are clearly beyond what one can expect a systematic modelling of irrelevant corrections to achieve \cite{Slevin1999b,Rodriguez2008MultifractalAveraging}. 
Nevertheless, using the crossings defined by considering just system sizes $M^2=6^2$ and $M^2=8^2$ from Fig.~\ref{fig:LocLength_Inverse_E0.4_2}(a), we find that the resulting ``phase boundary" faithfully follows the trend for the contours of DOS and $\langle r \rangle$ values as shown in Fig.~\ref{fig:dos_r_l31}(b).
%
Similarly, the $\langle r \rangle$ values reach $\langle r \rangle_\text{Sur}$ when $W \gtrsim 6$. For $W \lesssim 2$, the truly localized $\langle r \rangle_\text{Poi}$ ($\sim 0.38$) is not attained, but at least we find that $\langle r \rangle$ drops significantly to $\sim 0.45$. Again as in the case of the TMM data, no clear, system size-independent transition point emerges for the system sizes studied by us. 

In summary, the results at $E=0.4$ indicate the presence of a non-conventional ``inverse transition"al change from localized to extended regime as $W$ increases close to the macroscopic degeneracy of CLS. This seems similar to the proposed ``inverse" transition reported in a 3D all-band-flat network in the regime of weak uncorrelated disorder~\cite{Goda2006a,Nishino2007}. 

\subsection{The Lieb lattice $\mathcal{L}_3(2)$ and beyond}\label{sec:L32}

\begin{figure*}[tbh]
    \centering
    (a)\includegraphics[width=0.99\columnwidth]{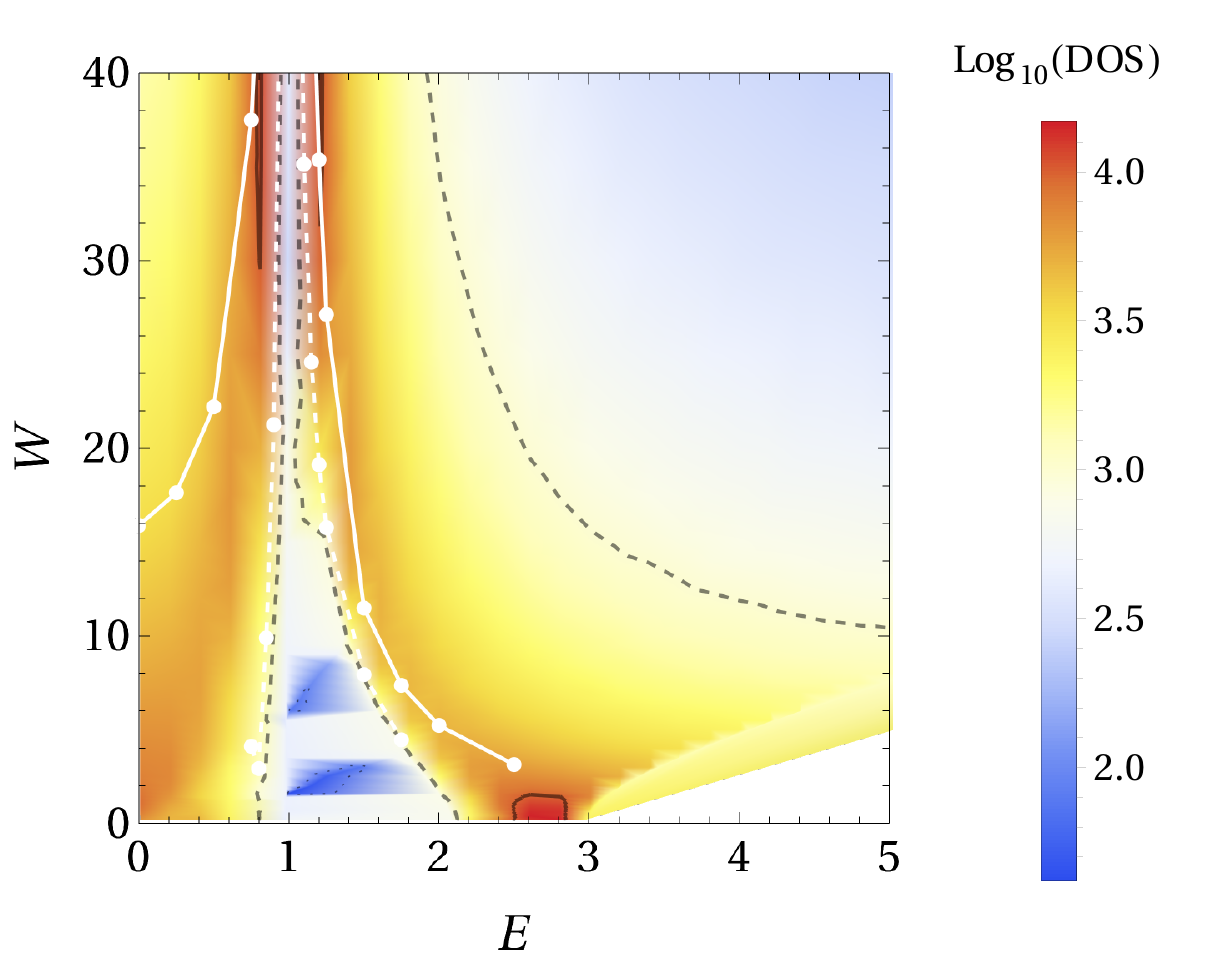}
    (b)\includegraphics[width=0.88\columnwidth]{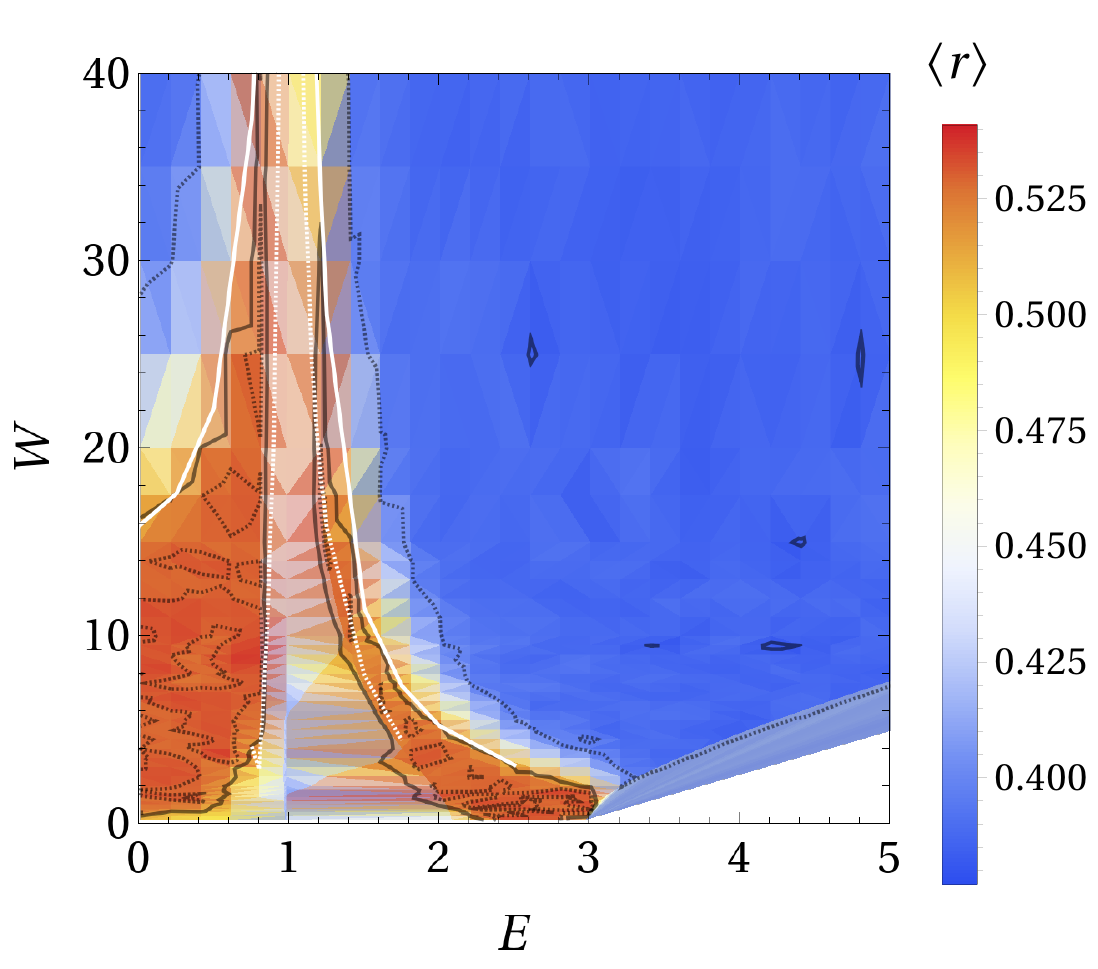}
    \caption{
    Energy $E$ and disorder $W$ dependant (a) DOS and (b) $r$-values for $\mathcal{L}_{3}(2)$. Both panels have been computed with the same parameters as in Fig.\ \ref{fig:dos_r_l31}, except that the minimal energy spacing increases to $\Delta E=0.2$, giving in total $1581$ individual $(E,W)$ pairs.
    The flat-band states at $E=1$ are not shown in both panels for clarity.
    The dark lines are as in Fig.\ \ref{fig:dos_r_l31} in (a) given by $10^3$ (dashed) and $10^4$ (solid) states, while for (b) they correspond to $\langle r\rangle=0.53$ (dashed), $0.5145$ (solid) in the red region, $\langle r\rangle=0.4$ (dashed) and $0.38$ (solid) in the blue region.
%
    Also as in Fig.\ \ref{fig:dos_r_l31}, the white lines in (a) and (b) denote estimates of the phase boundaries obtained by small-$M$ TMM with the $2$ different lines corresponding to the crossings of $\Lambda_M$ values between $M=6$ and $8$ from localized-to-delocalized (solid) and delocalized-to-localized (dashed) 
    behaviour upon decreasing $W$ at constant $E$. 
    }
    \label{fig:dos_r_l32}
\end{figure*}

We now briefly sketch the situation for the other Lieb lattices $\mathcal{L}_3(2)$, $\mathcal{L}_3(3)$  and $\mathcal{L}_3(4)$. 
For $\mathcal{L}_3(2)$, we show DOS, $\langle r \rangle$-based phase diagram and TMM-based approximate phase boundaries in Fig.\ \ref{fig:dos_r_l32}. 
The CLS at $E\pm 1$ are not explicitly shown in the figure 
but clearly visible by the behaviour of the non-CLS states around them. 
There is an identical signature of depletion of states, as for $\mathcal{L}_3(1)$, in the small $W$ region when $E \rightarrow 1^{\pm}$. On the other hand, for both large $E$ and $W$ the DOS depletes and the $\langle r \rangle$-values indicate localized behavior. Two extended regions emerge, both of which tend to lie close to the region of the CLS when $W \gtrsim 20$. These results are supported again from estimates based on TMM for $M^2=6^2$ and $M^2=8^2$. We note that due to the absence of CLS for $E=0$, we can indeed observe the usual change from extended to localized behaviour upon increasing $W$ with $W_c\sim 16$ marking the boundary between both regimes.
We can also find the ``inverse" behaviour again, e.g.\ for $\mathcal{L}_3(2)$ for $1 \lesssim E \lesssim 2$, where increasing $W$ leads to a change from localized to delocalized behaviour.





This trend continues for $\mathcal{L}_3(3)$  and $\mathcal{L}_3(4)$ (cp.\ supplemental material \cite{supp} and Figures therein): the originally dispersive bands, when $W=0$, move their states closer to the CLS upon increasing $W$, reducing the DOS for energies further from the CLS-energies and eventually localizing these. A sizable part of the spectrum moves closer to the CLS-energies, with states being moved onto the Lieb sites as shown for $\mathcal{L}_3(1)$.

\section{Conclusions}
\label{sec:conclusions}


As expected, the disorder $\varepsilon^{c}_\mathbf{X}$, together with the \emph{order} $\varepsilon^{L}_\mathbf{X}=0$, retains the distinction between CLS and the rest of the states, leaving the CLS are unchanged for any $W$. 
The converse is manifestly not the case: about half of the non-CLS states for, e.g., ${\cal L}_3(1)$ get pushed in energy close to the energy of the CLS and become evermore concentrated on the Lieb sites.
This leads to an accumulation of DOS near the CLS energies and, ultimately, to the existence of seemingly extended states even for very strong $W$ for all the ${\cal L}_3(1)$ to ${\cal L}_3(4)$ probed here. Indeed, for the system sizes studied, we cannot identify an upper critical disorder strength $W_\text{max}$ such that all states would be localized beyond this value. Instead, we find that the mobility edges are pushed towards large $W$ values, much larger than what is commonly observed for a regular cubic Anderson lattice \cite{Rodriguez2011MultifractalTransition}.


For the transition from extended to localized behaviour upon increasing $E$ or $W$ in the phase diagrams we find that the critical properties can be extracted as usual via FSS with critical exponent $\nu$ compatible with the usual value of the cubic Anderson lattice \cite{Slevin1999b,Rodriguez2011MultifractalTransition}. Hence, although the changes to the phase diagrams are drastic, the universal nature of the transition at this phase boundary does not change.
However, when instead decreasing $E$ and $W$ from the extended regimes, we do not see a clear signature of a transition as function of a single critical parameter strength. Rather, it appears that the changes of phase behaviour do no follow traditional scaling or require much larger system sizes to reach the scaling regime.

Overall, the model presents a situation where upon increasing $W$, the CLS are retained while non-CLS states are forced to become more and more CLS-like, in terms of energy as well as in terms of spatial location. As mentioned in the Introduction, CLS states are among a class of states that might become relevant for future information storage devices. Our result hence suggest a way in which disorder is not detrimental to such an application, but rather enhances the stability of the CLS. 
%
While solid-state devices with the chosen highly-correlated disorder/order distribution appear unlikely to become readily available soon, a much simpler route could be via cold atoms in optical lattices \cite{Shen2010,Goldman2011,Apaja2010} or in photonic band-gap systems \cite{Mukherjee2015a,Vicencio2015a,Guzman-Silva2014,Diebel2016,Taie2015,Nixon2013}
where single-site potential modulation has become routine \cite{Leykam2018c}. In such experimental and hence finite set-ups, it may be that the relevance of our finite size results is even more important than any large scale limit.
Last, it should be clear that an investigation of the influence on many-body interaction, in the presence of the CLS-preserving disorder considered here, should be most insightful. 

\section*{\label{sec:acknow}Acknowledgments}
We gratefully acknowledge the National Natural Science Foundation of China (Grant No.\ 11874316), the National Basic Research Program of China (2015CB921103), the Program for Changjiang Scholars and Innovative Research Team in University (Grant No.\ IRT13093), the Furong Scholar Program of Hunan Provincial Government (R.A.R.) for financial support, the Postgraduate Scientific Research Innovation Project of Hunan Province (No.\ CX20210515) and the Xiangtan University Graduate Research Innovation Project (No.\ XDCX2021B123). 
We thank Warwick's Scientific Computing Research Technology Platform for computing time and support. UK research data statement: Data accompanying this publication are available from \cite{Liu2022DataLattices}.



%
\bibliographystyle{prsty}

\begin{thebibliography}{}

\end{thebibliography}


\begin{thebibliography}{10}

\bibitem{Anderson1958c}
P.~W. Anderson, Physical Review {\bf 109},  1492  (1958).

\bibitem{Krameri1993}
B. Kramer and A. MacKinnon, Reports on Progress in Physics {\bf 56},  1469
  (1993).

\bibitem{2003AndersonRamifications}
{\em {Anderson Localization and Its Ramifications}}, Vol.~630 of {\em Lecture
  Notes in Physics}, edited by T. Brandes and S. Kettemann (Springer Berlin
  Heidelberg, Berlin, Heidelberg, 2003), p.\ 630.

\bibitem{Evers2008}
F. Evers and A.~D. Mirlin, Reviews of Modern Physics {\bf 80},  1355  (2008).

\bibitem{Abrahams1979ScalingDimensions}
E. Abrahams, P.~W. Anderson, D.~C. Licciardello, and T.~V. Ramakrishnan,
  Physical Review Letters {\bf 42},  673  (1979).

\bibitem{Bulka1985}
B. Bulka, B. Kramer, and A. MacKinnon, Z. Phys. B -Condensed Matter {\bf 60},
  13  (1985).

\bibitem{Aubry1980AnalyticityLattices}
S. Aubry and G. Andr{\'{e}}, {Analyticity breaking and Anderson localization in
  incommensurate lattices}, 1980.

\bibitem{Izrailev1999}
F.~M. Izrailev and A.~A. Krokhin, Physical Review Letters {\bf 82},  4062
  (1999).

\bibitem{Derzhko2015a}
O. Derzhko, J. Richter, and M. Maksymenko, International Journal of Modern
  Physics B {\bf 29},  1530007  (2015).

\bibitem{Leykam2018}
D. Leykam, A. Andreanov, and S. Flach, Advances in Physics: X {\bf 3},  1473052
   (2018).

\bibitem{Leykam2018c}
D. Leykam and S. Flach, APL Photonics {\bf 3},  070901  (2018).

\bibitem{Sutherland1986b}
B. Sutherland, Physical Review B {\bf 34},  5208  (1986).

\bibitem{Lieb1989TwoModel}
E.~H. Lieb, Physical Review Letters {\bf 62},  1201  (1989).

\bibitem{Tang2011}
E. Tang, J.-W. Mei, and X.-G. Wen, Physical Review Letters {\bf 106},  236802
  (2011).

\bibitem{Neupert2011}
T. Neupert, L. Santos, C. Chamon, and C. Mudry, Physical Review Letters {\bf
  106},  236804  (2011).

\bibitem{Sun2011NearlyTopology}
K. Sun, Z. Gu, H. Katsura, and S. Das~Sarma, Physical Review Letters {\bf 106},
   236803  (2011).

\bibitem{Savary2017}
L. Savary and L. Balents, Reports on Progress in Physics {\bf 80},  016502
  (2017).

\bibitem{Balents2010}
L. Balents, Nature {\bf 464},  199  (2010).

\bibitem{Mielke1991a}
A. Mielke, Journal of Physics A: Mathematical and General {\bf 24},  L73
  (1991).

\bibitem{Tasaki1992a}
H. Tasaki, Physical Review Letters {\bf 69},  1608  (1992).

\bibitem{Mielke1993FerromagnetismModel}
A. Mielke and H. Tasaki, Communications in Mathematical Physics {\bf 158},  341
   (1993).

\bibitem{Ramirez1994}
A.~P. Ramirez, Annual Review of Materials Science {\bf 24},  453  (1994).

\bibitem{Danieli2020}
C. Danieli, A. Andreanov, and S. Flach, Physical Review B {\bf 102},  041116
  (2020).

\bibitem{Kuno2020a}
Y. Kuno, T. Orito, and I. Ichinose, New Journal of Physics {\bf 22},  013032
  (2020).

\bibitem{Miyahara2007BCSLattice}
S. Miyahara, S. Kusuta, and N. Furukawa, Physica C: Superconductivity {\bf
  460-462},  1145  (2007).

\bibitem{Julku2016}
A. Julku, S. Peotta, T.~I. Vanhala, D.-H. Kim, and P. T{\"{o}}rm{\"{a}},
  Physical Review Letters {\bf 117},  045303  (2016).

\bibitem{Kopnin2011}
N.~B. Kopnin, T.~T. Heikkil{\"{a}}, and G.~E. Volovik, Physical Review B {\bf
  83},  220503  (2011).

\bibitem{Peotta2015}
S. Peotta and P. T{\"{o}}rm{\"{a}}, Nature Communications {\bf 6},  8944
  (2015).

\bibitem{Tovmasyan2018PreformedBands}
M. Tovmasyan, S. Peotta, L. Liang, P. T{\"{o}}rm{\"{a}}, and S.~D. Huber,
  Physical Review B {\bf 98},  134513  (2018).

\bibitem{Mondaini2018}
R. Mondaini, G.~G. Batrouni, and B. Gr{\'{e}}maud, Physical Review B {\bf 98},
  155142  (2018).

\bibitem{Aoki2020TheoreticalSuperconductivity}
H. Aoki, Journal of Superconductivity and Novel Magnetism {\bf 33},  2341
  (2020).

\bibitem{Abilio1999}
C.~C. Abilio, P. Butaud, T. Fournier, B. Pannetier, J. Vidal, S. Tedesco, and
  B. Dalzotto, Physical Review Letters {\bf 83},  5102  (1999).

\bibitem{Shen2010}
R. Shen, L.~B. Shao, B. Wang, and D.~Y. Xing, Physical Review B {\bf 81},
  041410  (2010).

\bibitem{Goldman2011}
N. Goldman, D.~F. Urban, and D. Bercioux, Physical Review A {\bf 83},  063601
  (2011).

\bibitem{Apaja2010}
V. Apaja, M. Hyrk{\"{a}}s, and M. Manninen, Physical Review A {\bf 82},  041402
   (2010).

\bibitem{Mukherjee2015a}
S. Mukherjee, A. Spracklen, D. Choudhury, N. Goldman, P. {\"{O}}hberg, E.
  Andersson, and R.~R. Thomson, Physical Review Letters {\bf 114},  245504
  (2015).

\bibitem{Vicencio2015a}
R.~A. Vicencio, C. Cantillano, L. Morales-Inostroza, B. Real, C.
  Mej{\'{i}}a-Cort{\'{e}}s, S. Weimann, A. Szameit, and M.~I. Molina, Physical
  Review Letters {\bf 114},  245503  (2015).

\bibitem{Guzman-Silva2014}
D. Guzm{\'{a}}n-Silva, C. Mej{\'{i}}a-Cort{\'{e}}s, M.~A. Bandres, M.~C.
  Rechtsman, S. Weimann, S. Nolte, M. Segev, A. Szameit, and R.~A. Vicencio,
  New Journal of Physics {\bf 16},  063061  (2014).

\bibitem{Diebel2016}
F. Diebel, D. Leykam, S. Kroesen, C. Denz, and A.~S. Desyatnikov, Physical
  Review Letters {\bf 116},  183902  (2016).

\bibitem{Taie2015}
S. Taie, H. Ozawa, T. Ichinose, T. Nishio, S. Nakajima, and Y. Takahashi,
  Science Advances {\bf 1},  e1500854  (2015).

\bibitem{Nixon2013}
M. Nixon, E. Ronen, A.~A. Friesem, and N. Davidson, Physical Review Letters
  {\bf 110},  184102  (2013).

\bibitem{Rontgen2019}
M. R{\"{o}}ntgen, C.~V. Morfonios, I. Brouzos, F.~K. Diakonos, and P.
  Schmelcher, Physical Review Letters {\bf 123},  080504  (2019).

\bibitem{Chalker2010a}
J.~T. Chalker, T.~S. Pickles, and P. Shukla, Physical Review B {\bf 82},
  104209  (2010).

\bibitem{Leykam2013}
D. Leykam, S. Flach, O. Bahat-Treidel, and A.~S. Desyatnikov, Physical Review B
  {\bf 88},  224203  (2013).

\bibitem{Flach2014a}
S. Flach, D. Leykam, J.~D. Bodyfelt, P. Matthies, and A.~S. Desyatnikov, EPL
  (Europhysics Letters) {\bf 105},  30001  (2014).

\bibitem{Leykam2017}
D. Leykam, J.~D. Bodyfelt, A.~S. Desyatnikov, and S. Flach, The European
  Physical Journal B {\bf 90},  1  (2017).

\bibitem{Bilitewski2018}
T. Bilitewski and R. Moessner, Physical Review B {\bf 98},  1  (2018).

\bibitem{Shukla2018a}
P. Shukla, Physical Review B {\bf 98},  054206  (2018).

\bibitem{Mao2020b}
X. Mao, J. Liu, J. Zhong, and R.~A. R{\"{o}}mer, Physica E: Low-Dimensional
  Systems and Nanostructures {\bf 124},  114340  (2020).

\bibitem{Cadez2021}
T. {\v{C}}ade{\v{z}}, Y. Kim, A. Andreanov, and S. Flach, Physical Review B
  {\bf 104},  L180201  (2021).

\bibitem{Bodyfelt2014}
J.~D. Bodyfelt, D. Leykam, C. Danieli, X. Yu, and S. Flach, Physical Review
  Letters {\bf 113},  1  (2014).

\bibitem{Danieli2015}
C. Danieli, J.~D. Bodyfelt, and S. Flach, Physical Review B - Condensed Matter
  and Materials Physics {\bf 91},  1  (2015).

\bibitem{Liu2020a}
J. Liu, X. Mao, J. Zhong, and R.~A. R{\"{o}}mer, Physical Review B {\bf 102},
  174207  (2020).

\bibitem{supp} See Supplemental Material at [URL will be inserted by
  publisher] for (i) the construction of the CLS, (ii) further details on projected probabilities and participation ratios, (iii) results for $\mathcal{L}_3(3)$ and $\mathcal{L}_3(4)$ as well as (iv) the explicit resolution of the adaptive grids used for $\mathcal{L}_3(1)$ and $\mathcal{L}_3(2)$.

\bibitem{MacKinnon1983a}
A. MacKinnon and B. Kramer, Zeitschrift f{\"{u}}r Physik B Condensed Matter
  {\bf 53},  1  (1983).

\bibitem{Note2}
We stress that true convergence has to be computed via error statistics of the
  \protect \emph {changes} in the $1/\lambda _M$ \cite
  {MacKinnon1983a,Liu2020a}.

\bibitem{Romer2022NumericalLocalization}
R.~A. R{\"{o}}mer, {Numerical methods for localization}, 2022.

\bibitem{Belitz1994a}
D. Belitz and T.~R. Kirkpatrick, Reviews of Modern Physics {\bf 66},  261
  (1994).

\bibitem{Slevin1999b}
K. Slevin and T. Ohtsuki, Physical Review Letters {\bf 82},  382  (1999).

\bibitem{Rodriguez2011MultifractalTransition}
A. Rodriguez, L.~J. Vasquez, K. Slevin, and R.~A. R{\"{o}}mer, Physical Review
  B {\bf 84},  134209  (2011).

\bibitem{Note3}
We call {\protect \sc LaPack} function DSYEV() from the 2022 Math Kernel
  Library (MKL) to calculate eigenvalues and eigenvectors.

\bibitem{Bollhofer2007JADAMILU:Matrices}
M. Bollh{\"{o}}fer and Y. Notay, Computer Physics Communications {\bf 177},
  951  (2007).

\bibitem{Oganesyan2007c}
V. Oganesyan and D.~A. Huse, Physical Review B {\bf 75},  155111  (2007).

\bibitem{Atas2013b}
Y.~Y. Atas, E. Bogomolny, O. Giraud, and G. Roux, Physical Review Letters {\bf
  110},  084101  (2013).

\bibitem{Sa2020}
L. S{\'{a}}, P. Ribeiro, and T. Prosen, Physical Review X {\bf 10},  021019
  (2020).

\bibitem{Luo2021a}
X. Luo, T. Ohtsuki, and R. Shindou, Physical Review Letters {\bf 126},  090402
  (2021).

\bibitem{Note4}
These values have been computed from $1000$ independent random realization of
  $20000 \times 20000$ matrices for GOE and $1000$ realization of random
  diagonals with $20000$ entries, respectively. The calculations also yield
  $\langle r \rangle _\protect \text {GOE} = 0.5307(1)$ which is in excellent
  agreement with the best fit value obtained by Atas et al.\ \cite {Atas2013b}
  for $\langle r \rangle _\protect \text {Sur}$.

\bibitem{Note5}
We note that the estimated $W_c\approx 40.2$ at energy $E=0.4$ is much higher
  also than the largest transition measured at $E=0$ in the uniformly
  disordered $\protect \mathcal {L}_3(1)$ reported in Ref.~\cite {Liu2020a}.

\bibitem{Ramachandran2017}
A. Ramachandran, A. Andreanov, and S. Flach, Physical Review B {\bf 96},
  161104  (2017).

\bibitem{Rodriguez2008MultifractalAveraging}
A. Rodriguez, L.~J. Vasquez, and R.~A. R{\"{o}}mer, Physical Review B {\bf 78},
   195107  (2008).

\bibitem{Goda2006a}
M. Goda, S. Nishino, and H. Matsuda, Physical Review Letters {\bf 96},  126401
  (2006).

\bibitem{Nishino2007}
S. Nishino, H. Matsuda, and M. Goda, Journal of the Physical Society of Japan
  {\bf 76},  024709  (2007).

\bibitem{Liu2022DataLattices}
J. Liu, C. Danieli, J. Zhong, and R.~A. R{\"{o}}mer, {Data for "Unconventional
  delocalization in three-dimensional Lieb lattices"}, 2022, url:http://wrap.warwick.ac.uk/169105/. 

\end{thebibliography}


\ifNOSUP\end{document}\else%

\clearpage\newpage
\setcounter{section}{0}
\setcounter{figure}{0}
\setcounter{table}{0}
\def\thesection{S\arabic{section}}
\def\thefigure{S\arabic{figure}}
\def\thetable{S\arabic{table}}
\setcounter{page}{1}
\pagestyle{plain}

\section*{Supplemental Material}

{\center
\textbf{Unconventional delocalization in three-dimensional Lieb lattices}\\[2ex]

\noindent%
Jie Liu$^{1}$, Carlo Danieli$^{2}$, Jianxin Zhong$^{1}$, Rudolf A R\"{o}mer$^{1,3}$\\[2ex]

$^1${School of Physics and Optoelectronics, Xiangtan University, Xiangtan 411105, China}\\
$^2${Max Planck Institute for the Physics of Complex Systems, Dresden D-01187, Germany}\\
$^3${Department of Physics, University of Warwick, Coventry, CV4 7AL, United Kingdom}\\
}
	
\hspace*{2ex}

\section{Construction of CLS in $\mathcal{L}_3(n)$}\label{appA}

Each case of the generalized Lieb lattice $\mathcal{L}_3(n)$ for $n=1$ to $4$ discussed in this work (Fig.~\ref{fig:Lieb_schematic} of the main text) possesses \revision{ $2n$ flat bands whose irreducible CLS have strictly non-zero amplitude in the Lieb sites only of 2D plaquettes of the lattice.}  

\begin{figure}[tb]
    \centering
    \includegraphics[width=\columnwidth]{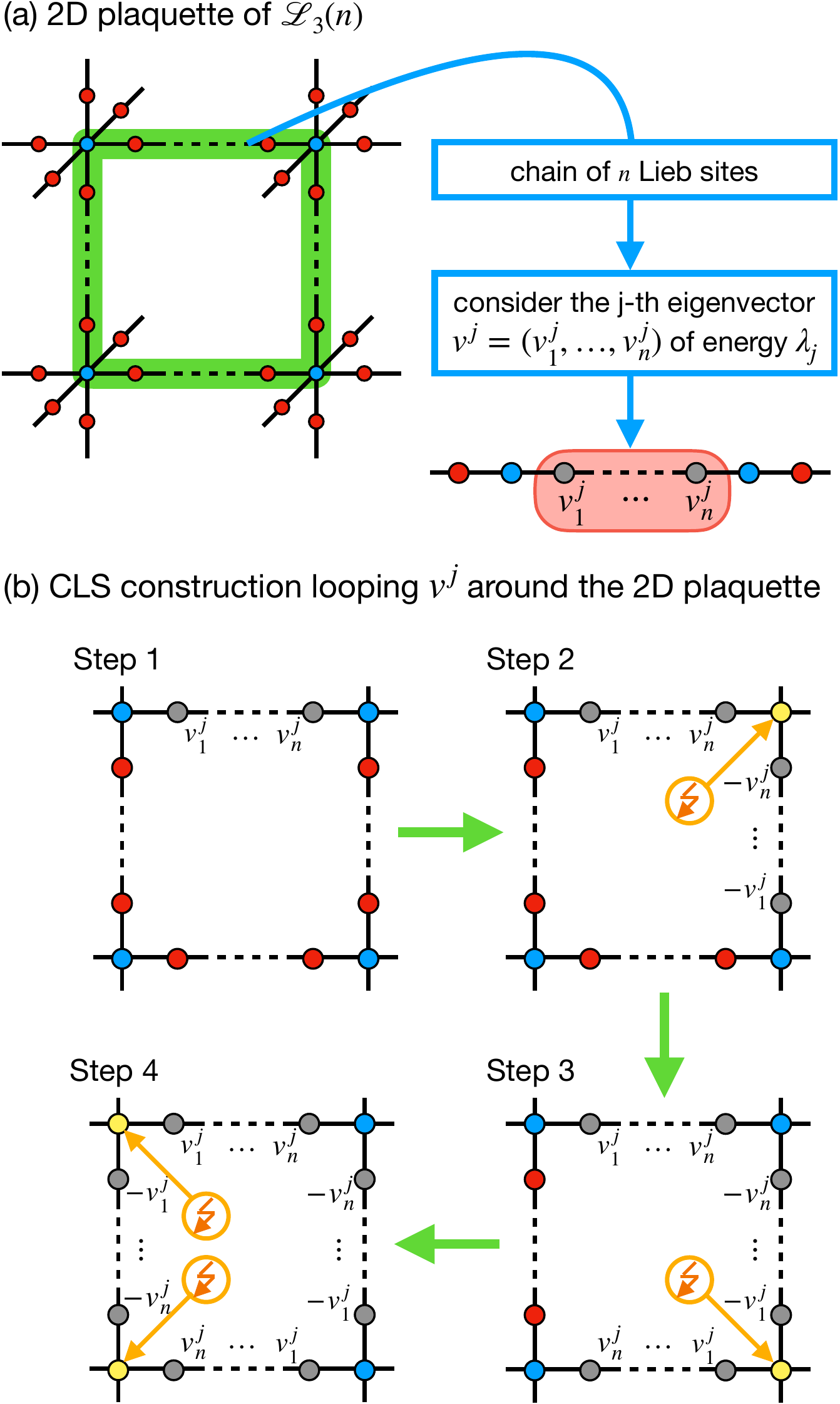} 
    \caption{
    (a) Section of the lattice $\mathcal{L}_3(n)$ around a 2D plaquette (green color). We consider the 1D chain formed by the  $n$ Lieb sites sandwiched between two cube sites along one edge of the plaquette, and highlight one of the eigenvector $v^j$ of eigenenergy $\lambda_j$. 
    (b) Four steps construction of a CLS of  $\mathcal{L}_3(n)$ with energy $\lambda_j$ by looping $v^j$ around the 2D plaquette. In each step, we highlight in yellow and with the thunder symbol the cube site where the destructive interference is enforced. 
    }
    \label{fig:app_CLS_n}
\end{figure}

Thus, for a generic lattice $\mathcal{L}_3(n)$, it is instructive to focus on a 2D plaquette enclosed within four neighbouring cube sites -- highlighted in green color in Fig.~\ref{fig:app_CLS_n}(a), left side. We consider the 1D chain formed by the $n$ Lieb sites in one of the four edges of the plaquette. 
We then denote by $v^j = (v_1^j,\dots,v_n^j)$ one of the $n$ eigenvectors of this 1D chain with eigenenergy $\lambda_j$ -- \revision{as sketched in Fig.~\ref{fig:app_CLS_n}(a)}.


A CLS of $\mathcal{L}_3(n)$ is then constructed by looping $v^j$ around the plaquette as shown in  Fig.~\ref{fig:app_CLS_n}(b). 
Proceeding clockwise from top-left, the construction works as follows: \\
-- in \underline{Step 1} we consider $v^j$ along the upper edge. Hence, the amplitudes on the Lieb sites next to the top-left and top-right cube sites are $v_1^j$ and $v_n^j$, respectively; \\
-- in \underline{Step 2} we ensure destructive interference in the top-right cube site of the plaquette (colored in yellow and indicated with a thunder symbol) by setting $- v_n^j$ in the top Lieb site of the right edge of the plaquette. The remaining $n-1$ Lieb sites are then filled by $- v_i^j$ with the index $i$ running in a downward direction. The amplitude of the bottom Lieb site of this chain thus is $- v_1^j$; \\ 
-- in \underline{Step 3} we ensure destructive interference in the bottom-right cube site of the plaquette (highlighted alike Step 2) by setting $ v_1^j$ in the right Lieb site of the bottom edge of the plaquette.
The remaining $n-1$ Lieb sites are then filled by $ v_i^j$ with the index $i$ running in a leftward direction. The amplitude of the most-left Lieb site of this chain thus is $v_n^j$; \\
-- in \underline{Step 4} we ensure destructive interference in the highlighted bottom-left cube site of the plaquette by setting $ - v_n^j$ in the bottom Lieb site of the left edge of the plaquette.
The remaining $n-1$ Lieb sites are then filled by  $- v_i^j$ with the index $i$ running  in a upward direction. The amplitude of the top Lieb site of this chain thus is $-v_1^j$, which ensures destructive interference in the highlighted top-left cube site. 

\revision{
This iterative procedure yields a compact eigenstate of energy $\lambda_j$. 
The $\mathcal{L}_3(n)$ posses a total of $3n+1$ Bloch bands, and 
in a cube cut-off of the lattice with $N$ unit-cells per side 
each band posses a total of $N^3$ states. 
Such cube version of $\mathcal{L}_3(n)$ hence support $3 N^3$ CLS -- {\it i.e.} as many as the number of 2D plaquettes in the cube. However, it easily follows that $N^3$ of these CLS are linear combination of the remaining compact states, yielding an irreducible number of $2 N^3$ CLS at energy $\lambda_j$. 
These states consequently form two flat bands for the $\mathcal{L}_3(n)$ at $E=\lambda_j$ -- or, equivalently, a double-counted flat band at $E=\lambda_j$.} 

\revision{
Since there exist $n$ eigenvector $\{v^i\}_{i=1}^n$ with energies $\lambda_i\neq\lambda_j$ for $i\neq j$ for the 1D chain of $n$ Lieb sites, this construction can be done $n$ times -- generating $n$ families of CLS for the lattice $\mathcal{L}_3(n)$, and therefore $2 n$ flat bands $\{\lambda_i\}_{i=1}^n$ -- or, equivalently, $n$ double-degenerate flat bands. }



\section{Projected probabilities beyond $E\geq 1.5$ and participation ratios}

\begin{figure*}[bth]
    \centering
    (a)\includegraphics[width=0.95\textwidth]{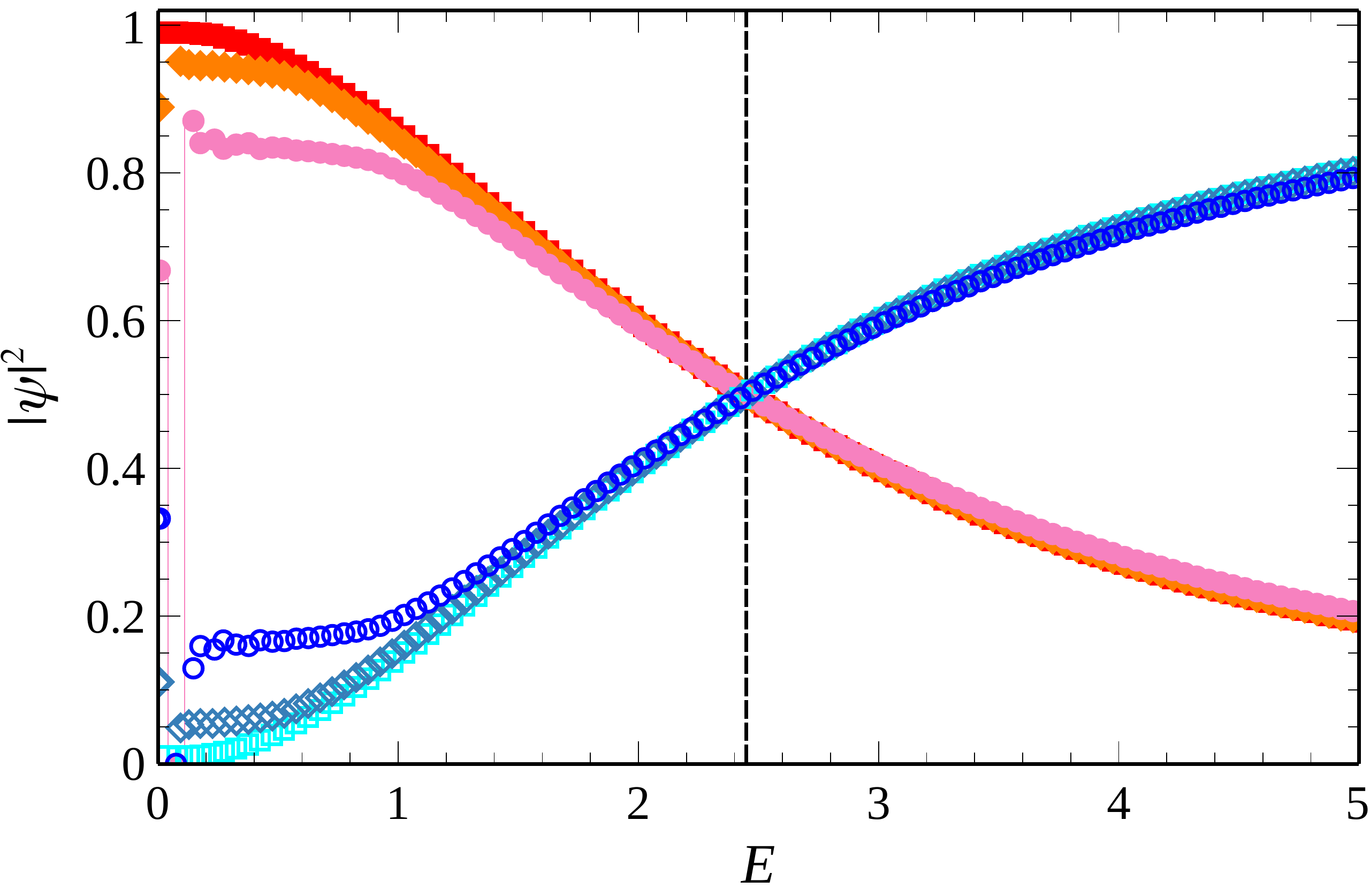}
    (b)\includegraphics[width=0.95\textwidth]{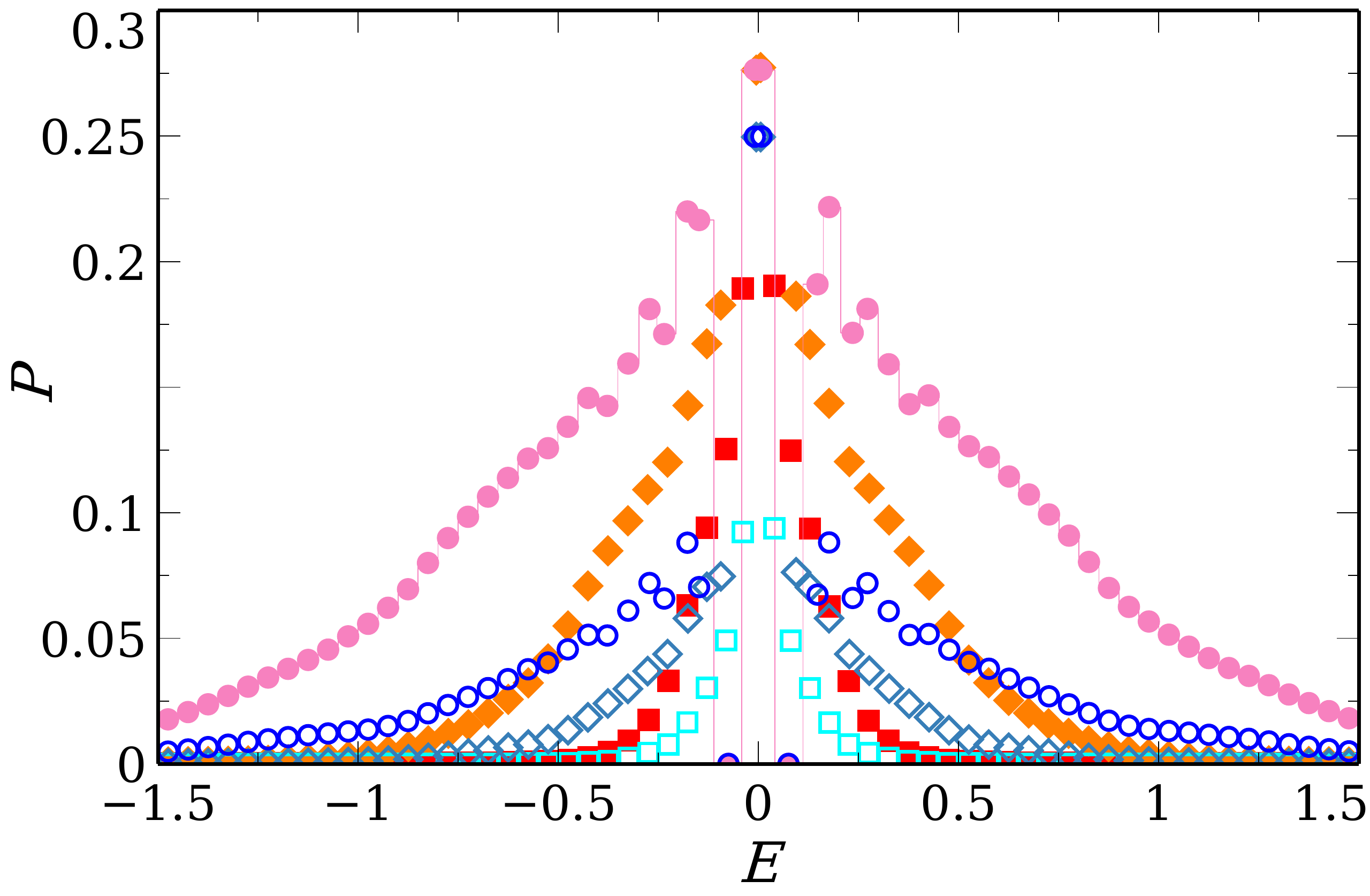}
    \caption{(a) Projected probabilities $|\psi(\vec{r})|^2$ for cube sites (blue colors, open symbols) and Lieb sites (red colors, filled symbols) with disorders $W=10$ ($\circ$), $20$ ($\Diamond$) and $50$ ($\Box$) for ${\cal L}_3(1)$ as in Fig.\ \ref{fig:probs_l31_2}, but with a different range in $E$ to highlight the crossing point at $E\sim 2.45$ (vertical dashed line).
    (b) Participation numbers expressed relative to the number of cube and Lieb sites. 
    In both panels, the line for Lieb sites with $W=10$ is given to highlight that the data points represent averages for $144$ potential configurations with energy resolution $\Delta E = 0.05$. The system size in all cases is $L= 4 \times 20^3$.}
    \label{fig:probs_l31_XL}
\end{figure*}

In Fig.\ \ref{fig:probs_l31_XL} we show the same data as in Fig.\ \ref{fig:probs_l31_2}, but now for $E\geq 0$. We find that the projected probabilities cross when $E\sim 2.45$. This happens at all $W$s that we have studied, \emph{i.e.} up to $W=100$. Closer investigation reveals that about $50\%$ of all $4\times N^3$ possible states are CLS, $25\%$ get shifted towards the CLS energies when $W$ increases, while the remaining $25\%$ localize and spread out into the full range of the spectrum. The relative participation numbers, \emph{i.e.} $P= 1/ \sum_{l=1}^{L_\text{max}} | \psi(\vec{r}_l)|^4 / (L_\text{max})$ as shown in panel (b) of Fig.\ \ref{fig:probs_l31_XL}, indicate that indeed appreciable $P$ are only observed close to the CLS energy $E=0$ for $\mathcal{L}_3(1)$. Note that here $L_\text{max}$ is the number of sites corresponding to cube and Lieb sites, \emph{i.e.} $N^3$ and $3n N^3$, respectively, for $\mathcal{L}_3(n)$.

\section{Results for Lieb lattices $\mathcal{L}_3(3)$ and $\mathcal{L}_3(4)$}

\begin{figure*}[tb]
    \centering
    $\mathcal{L}_{3}(3)$ \\
    (a)\includegraphics[width=0.95\columnwidth]{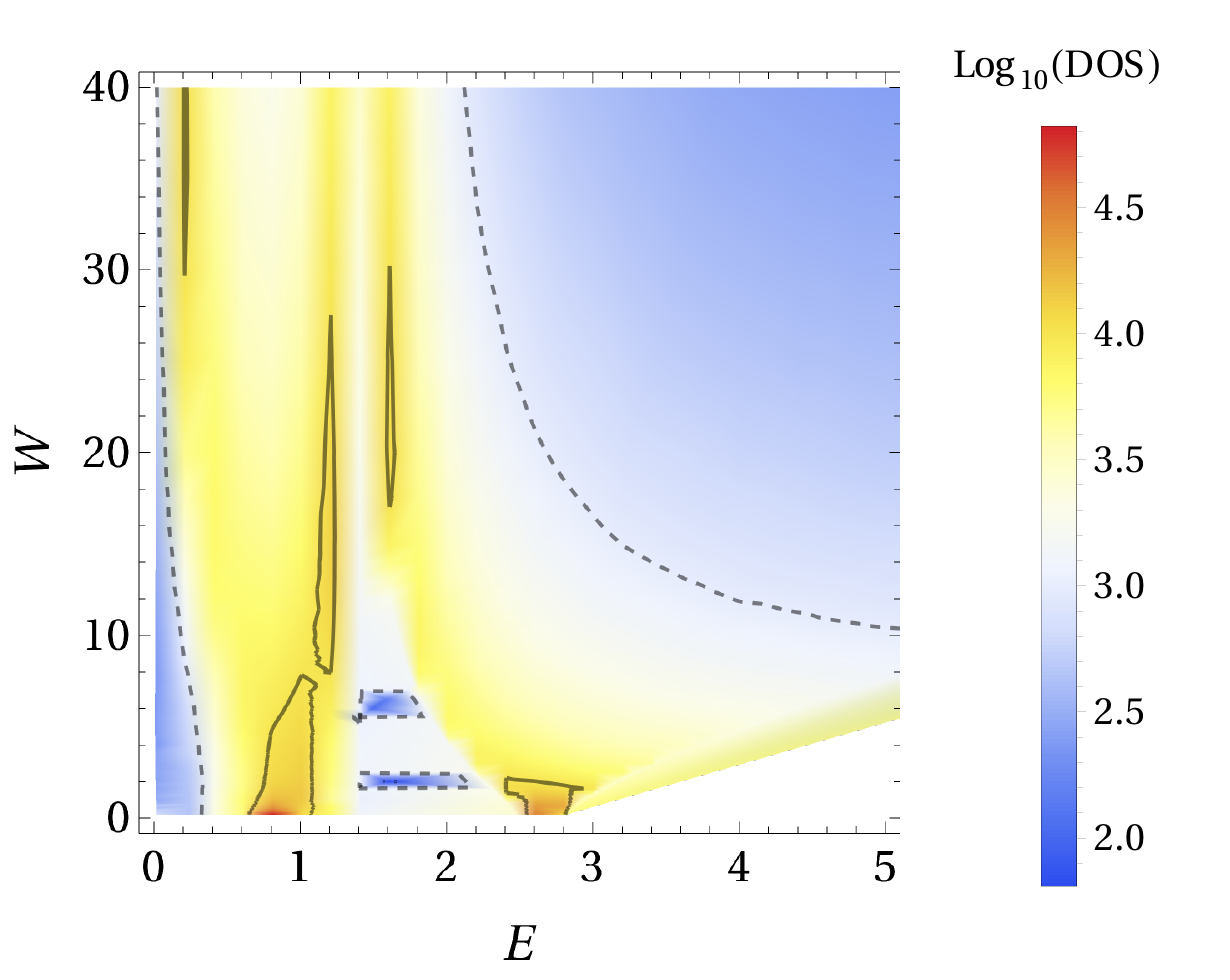}
    (b)\includegraphics[width=0.85\columnwidth]{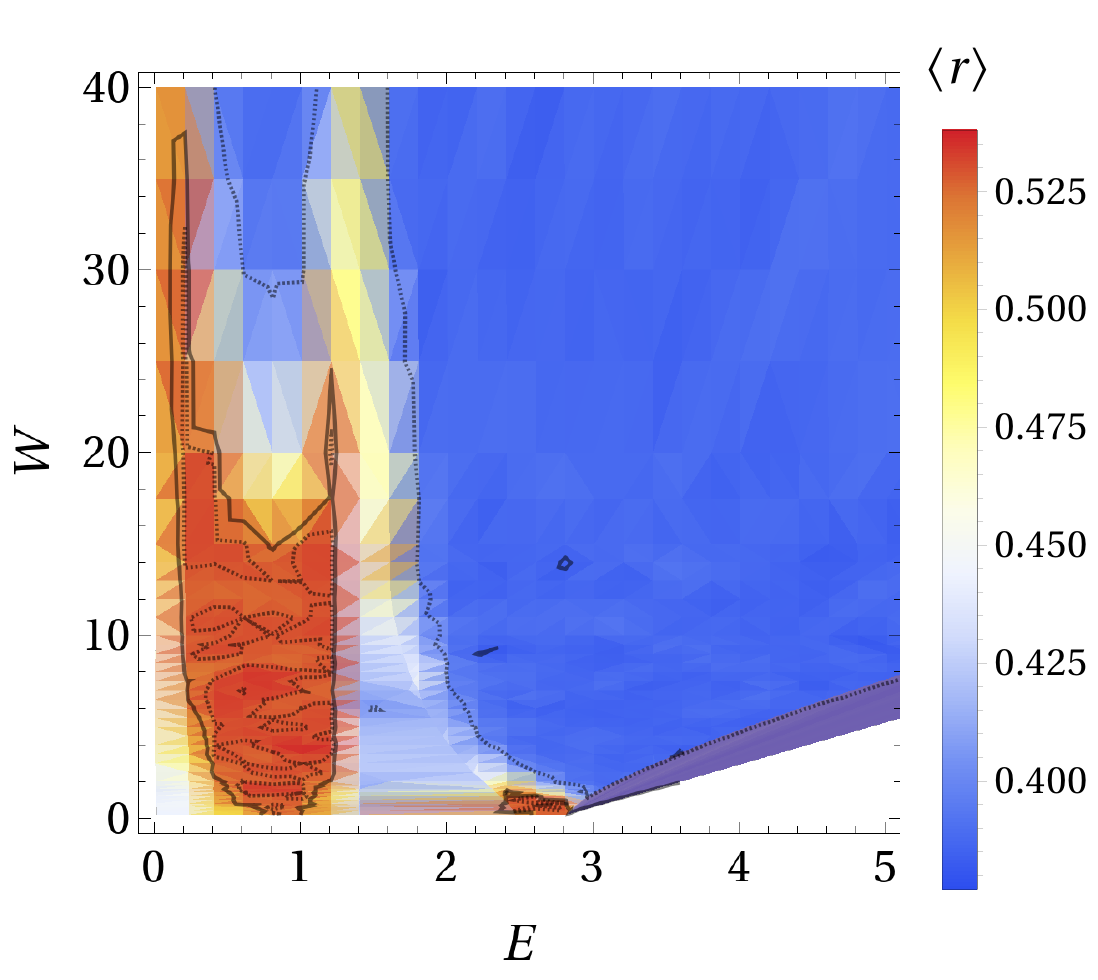}\\[2ex]
    $\mathcal{L}_{3}(4)$\\ 
    (c)\includegraphics[width=0.95\columnwidth]{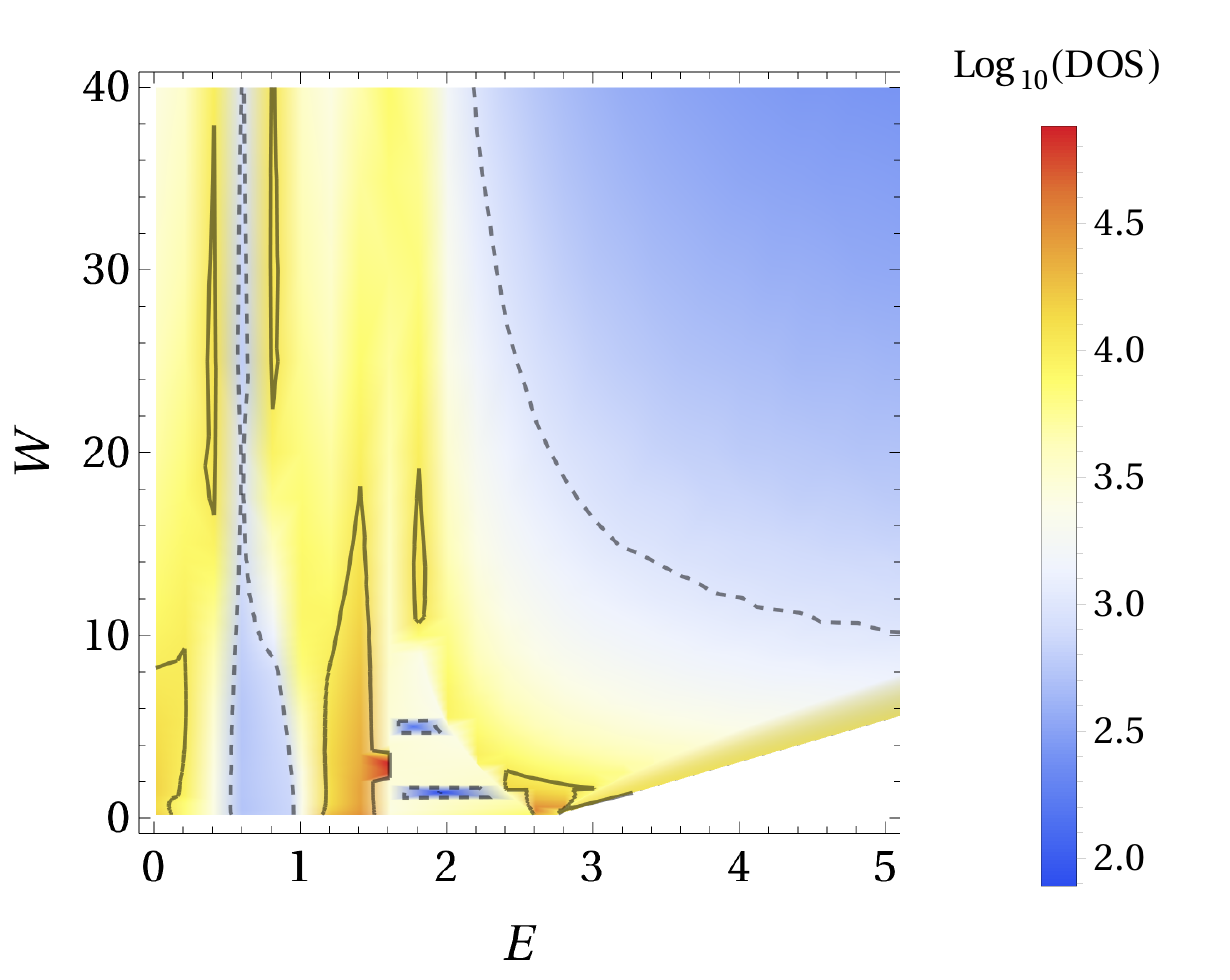}
    (d)\includegraphics[width=0.85\columnwidth]{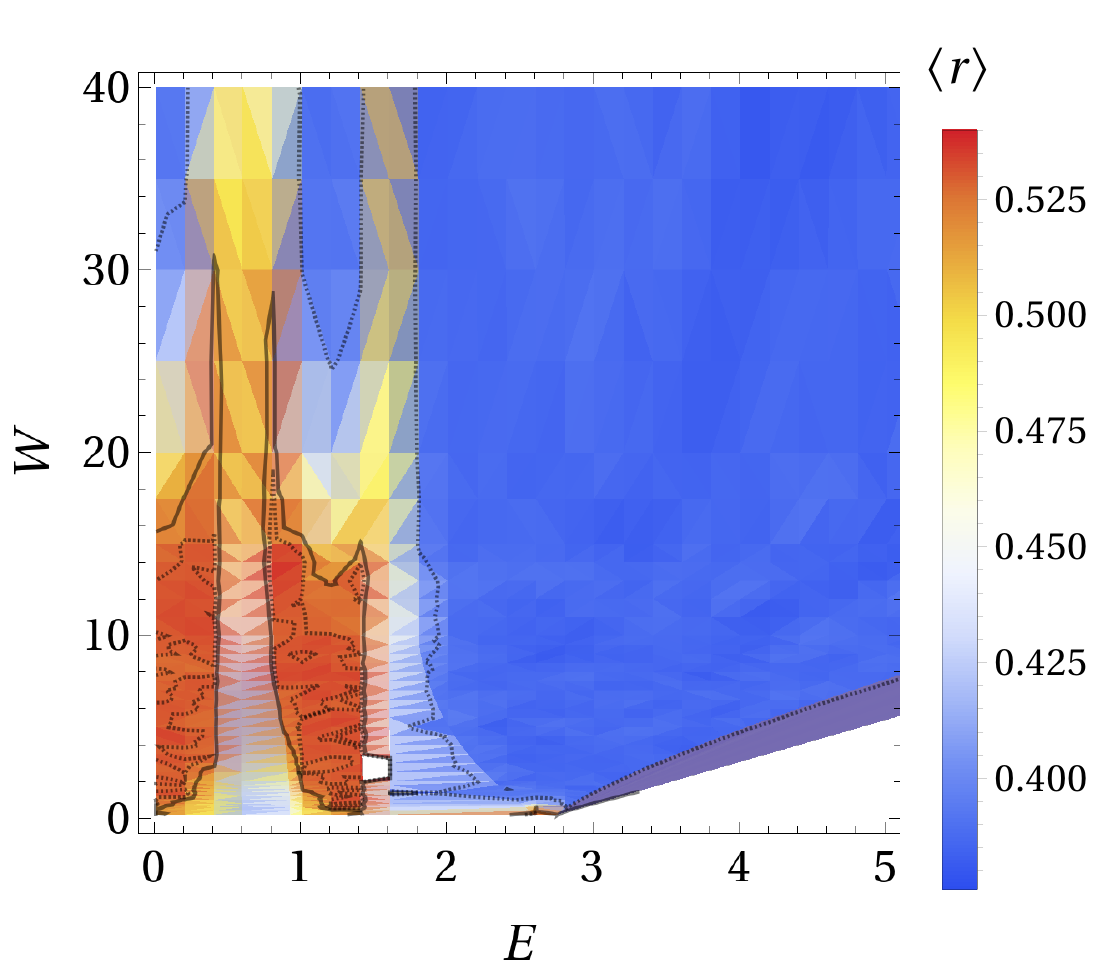}
    \caption{
    (a,c) DOS and (b,d) $r$-values for (a,b) $\mathcal{L}_{3}(3)$ and (c,d) $\mathcal{L}_{3}(4)$ similar to Fig.\ \ref{fig:dos_r_l32} with a total of $1587$ adaptive individual $(E,W)$ pairs.
    The flat-band states at (a,b) $E=0$ and $\sqrt{2}$ for $\mathcal{L}_{3}(3)$ and (c,d) at $E=(\sqrt{5}-1)/2 \sim 0.618$ and $(1+\sqrt{5})/2 \sim 1.618$ for $\mathcal{L}_{3}(4)$ are again not shown in all panels for clarity.
    The dark lines are also as before, \emph{i.e.} $10^3$ (dashed) and $10^4$ (solid) in (a,c), while they correspond to $\langle r\rangle=0.53$ (dashed), $0.5145$ (solid) in the red region, $\langle r\rangle=0.4$ (dashed) and $0.38$ (solid) in the blue region in (b,d).
%
%
     }
    \label{fig:dos_r_l34}
    \label{fig:dos_r_l33}
\end{figure*}

We plot DOS, the $\langle r \rangle$-based phase diagram and TMM-based approximate phase boundaries in Fig.\ \ref{fig:dos_r_l33} for $\mathcal{L}_3(3)$ and $\mathcal{L}_3(4)$.
The CLS at $E\pm 1$ are not explicitly indicated in the figures 
but clearly visible by the behaviour of the non-CLS states around them. 
There is an identical signature of depletion of states, as for $\mathcal{L}_3(1)$ and $\mathcal{L}_3(2)$, in the small $W$ regions when $E$ approaches the CLS energies.
For larger $W$, clear areas of localization behaviour emerge except for energies close to the CLS energies where even very strong disorder does not appear to suppress delocalized behaviour for the system sizes studied here. We can also find the ``inverse" behaviour again in various energy regions although better energy resolution would be needed to reproduce fine details such as given, e.g., for $\mathcal{L}_3(1)$ in Fig.\ \ref{fig:dos_r_l31}.\newline

\section{The $(E,W)$ grids used for $\mathcal{L}_3(1)$ and $\mathcal{L}_3(2)$}

\begin{figure*}[tb]
    \centering
    (a)\includegraphics[width=0.87\columnwidth]{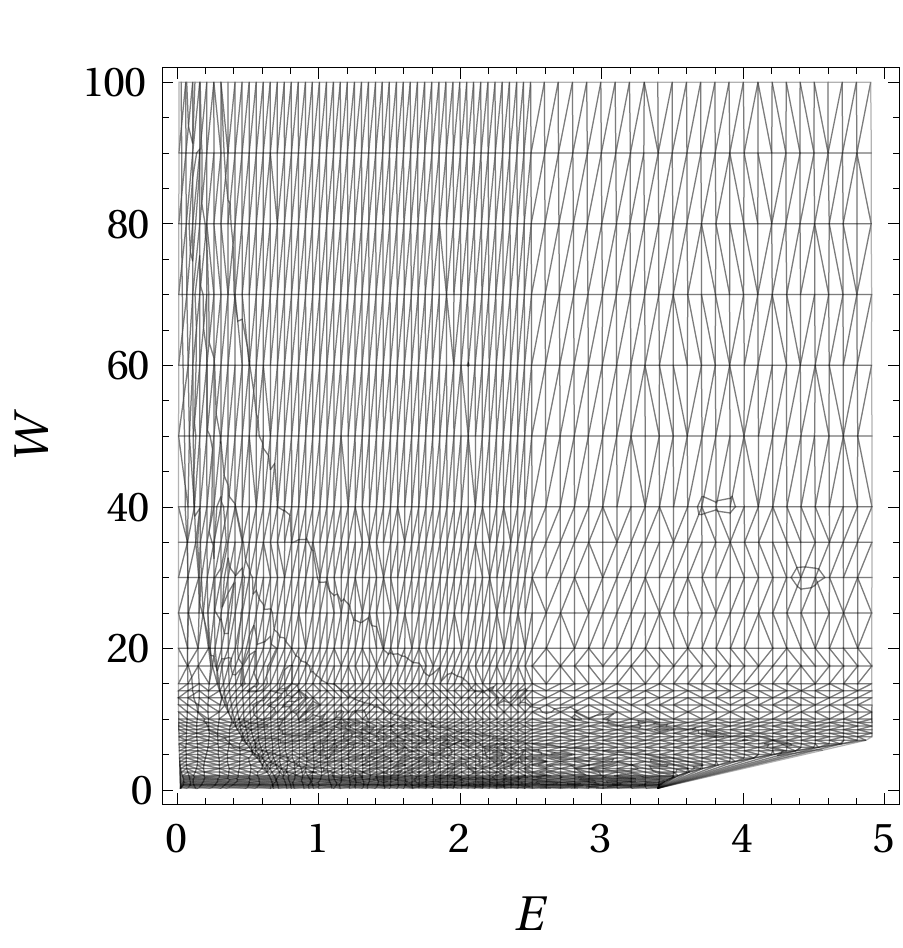}
    (b)\includegraphics[width=0.85\columnwidth]{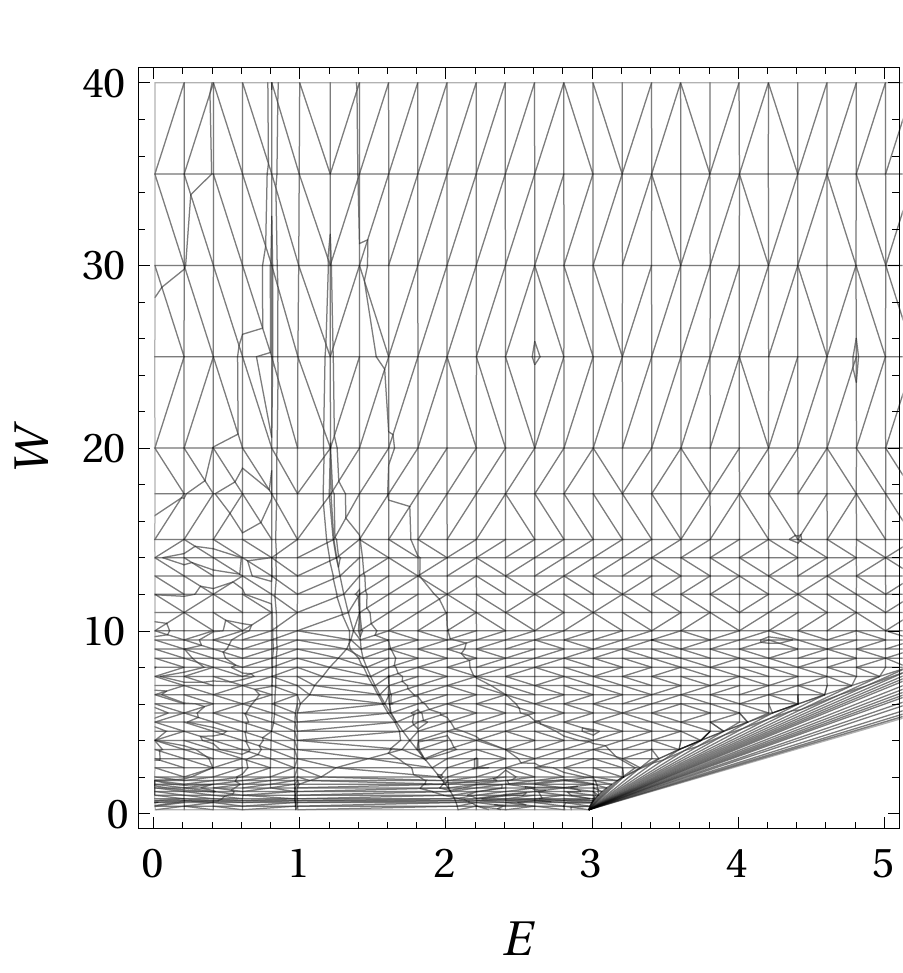}
    \caption{
    Representation of the underlying $(E,W)$ mesh that was used for DOS and $r$-values density and contour plots in (a) for $\mathcal{L}_{3}(1)$ and (b) for $\mathcal{L}_{3}(2)$ as shown in Figs. \ref{fig:dos_r_l31} and \ref{fig:dos_r_l32}.
    The contour lines for $\langle r\rangle=0.53$ (dashed), $0.5145$ (solid) in the delocalized region, and $\langle r\rangle=0.4$ (dashed) and $0.38$ (solid) in the localized region as plotted as before.
    The mesh resolution for $\mathcal{L}_{3}(3)$ and $\mathcal{L}_{3}(4)$ is comparable to the $\mathcal{L}_{3}(2)$ case shown here in (b). 
}
    \label{fig:mesh_dos_r_l3x}
\end{figure*}

In Fig.\ \ref{fig:mesh_dos_r_l3x}, we display the set of $(E,W)$ points that have been used to construct DOS and $\langle r \rangle$-value based phase diagrams in Figs.\ \ref{fig:dos_r_l31} and \ref{fig:dos_r_l32}. The non-regularity in the spacing of the $(E,W)$ points is due to the use of the sparse-matrix diagonalization routines. While we can give the routine a target energy, it is not guaranteed that such an energy exists in the spectrum. Hence we compute a mean energy $\langle E \rangle$ and use its value to anchor DOS and $\langle r \rangle$ values. This leads to the highly adaptive meshing structure presented in Fig.\ \ref{fig:mesh_dos_r_l3x}, which shows high resolution of $(\langle E\rangle,W)$ in areas with high DOS and much lower resolution for regions of low DOS. In this way, we can adaptively concentrate on regions with relevant data while also being able to reduce computational effort due to the use of the sparsity of the Hamiltonian matrix. 

\fi\end{document}